\documentclass[longauth]{aa} 
\usepackage{graphicx}
\usepackage{txfonts}
\usepackage{longtable}  
\usepackage{hyperref}
\hypersetup{
colorlinks=true,    
linkcolor=red,      
citecolor=blue,     
filecolor=magenta,  
urlcolor=black      
}
\usepackage{amsmath}	
\usepackage{amssymb}	
\usepackage{xtab,booktabs}   %
\usepackage{threeparttable}  

\usepackage{natbib,twoopt}
\bibpunct{(}{)}{;}{a}{}{,}             

\newcommand{\swift}{\textit{Swift}}

\makeatletter
\renewcommand*\aa@pageof{, page \thepage{} of \pageref*{LastPage}}
\makeatother

\begin{document}

   \title{
   GRB 180728A and SN 2018fip: the nearest high-energy cosmological gamma-ray burst with an associated supernova
\thanks{Based on observations
collected at the Very Large Telescope of the European Southern
Observatory,  Paranal, Chile (ESO programmes 
2101.D-5044, PI: A. Rossi; 0101.D-0648, PI: N. Tanvir}
   }
   \titlerunning{GRB 180728A and SN 2018fip}


      \author{
   A. Rossi\inst{1}\fnmsep\thanks{andrea.rossi@inaf.it}
   \and
   L. Izzo \inst{2,3}
   \and
   K. Maeda \inst{4}
   \and
   P. Schady \inst{5}
   \and
   D. B. Malesani \inst{6,7,8}
   \and
   D. A. Kann \inst{9,10}\thanks{Deceased}
   \and
   S. Klose \inst{11}
   \and 
   L. Amati \inst{1}
   \and
   P. D'Avanzo  \inst{12}
   \and
   A.~de Ugarte Postigo \inst{13,14}
   \and
   K.~E.~Heintz \inst{6,7} 
   \and
   A. Kumar \inst{15}
   \and 
   V. Lipunov  \inst{16} 
   \and 
   A. Martin-Carrillo  \inst{17}
   \and
   A. Melandri \inst{18}
   \and 
   A. M. Nicuesa Guelbenzu \inst{11}
   \and
   S.~R. Oates \inst{19}
   \and
   S. Schulze \inst{20}  
   \and
   J. Selsing \inst{3}  
   \and 
   R.~L.~C. Starling \inst{21}
   \and 
   G. Stratta  \inst{1}
   \and
   D. Vlasenko  \inst{16}
   \and
   P. Balanutsa  \inst{16}
   \and
   R. Brivio \inst{11}
   \and
   V. D'Elia \inst{22}
   \and
   B. Milvang-Jensen  \inst{6,7}
   \and
   E. Palazzi \inst{1}
   \and
   D.~A. Perley \inst{23} 
   \and
   A. Rau \inst{24}
   \and
   J. Sollerman \inst{25}
   \and
   N.~R. Tanvir \inst{21}
   \and
   T. Zafar  \inst{26}
           }
           
\institute{          
  INAF - Osservatorio di Astrofisica e Scienza dello Spazio, via Piero Gobetti 93/3, 40129 Bologna, Italy 
  \and INAF, Osservatorio Astronomico di Capodimonte, Salita Moiariello 16, 80131 Naples, Italy 
  \and DARK, Niels Bohr Institute, University of Copenhagen, Jagtvej 128, 2200 Copenhagen, Denmark 
  \and Department of Astronomy, Kyoto University, Kitashirakawa-Oiwake-cho, Sakyo-ku, Kyoto 606-8502, Japan 
  \and Department of Physics, University of Bath, Claverton Down, Bath BA2 7AY, UK 
  \and Cosmic Dawn Center (DAWN) 
  \and Niels Bohr Institute, University of Copenhagen, Jagtvej 128, 2200 Copenhagen N, Denmark 
  \and Department of Astrophysics/IMAPP, Radboud University, 6525 AJ Nijmegen, The Netherlands 
  \and Hessian Research Cluster ELEMENTS, Giersch Science Center, Max-von-Laue-Stra$\beta$e 12, Goethe University Frankfurt, Campus Riedberg, D-60438 Frankfurt am Main, Germany 
  \and Instituto de Astrof\'isica de Andaluc\'ia (IAA-CSIC), Glorieta de la Astronom\'ia s/n, 18008 Granada, Spain  
  \and Th\"uringer Landessternwarte Tautenburg, Sternwarte 5, 07778 Tautenburg, Germany 
  \and INAF - Osservatorio Astronomico di Brera, Via E. Bianchi 46, I-23807, Merate (LC), Italy 
  \and Artemis, Universit\'{e} de la C\^ote d'Azur, Observatoire de la C\^ote d'Azur, CNRS, 06304 Nice, France 
  \and Aix Marseille Univ, CNRS, LAM Marseille, France 
  \and Department of Physics, Royal Holloway - University of London, Egham, TW20 0EX, U.K. 
  \and Lomonosov Moscow State University, 119234 Moscow, Universitetskiy prospekt, 13  
  \and School of Physics and Centre for Space Research, University College Dublin, Belfield D04 V1W8, Dublin, Ireland 
  \and INAF, Osservatorio Astronomico di Roma, via Frascati 33, I-00078 Monte Porzio Catone (Roma), Italy. 
  \and Department of Physics, Lancaster University, Lancs LA1 4YB, UK 
  \and Center for Interdisciplinary Exploration and Research in Astrophysics (CIERA), Northwestern University, 1800 Sherman Avenue, Evanston, IL 60201, USA 
  \and School of Physics and Astronomy, University of Leicester, University Road, Leicester LE1 7RH, UK 
  \and Space Science Data Center (SSDC) - Agenzia Spaziale Italiana (ASI), Via del Politecnico snc, I-00133 Roma, Italy 
  \and Astrophysics Research Institute, Liverpool John Moores University, 146 Brownlow Hill, Liverpool L3 5RF, UK 
  \and Max-Planck-Institut f\"ur Extraterrestrische Physik, Giessenbachstra\ss{}e 1, 85748 Garching, Germany  
  \and The Oskar Klein Centre, Department of Astronomy, AlbaNova, SE-106 91 Stockholm , Sweden 
  \and School of Mathematical and Physical Sciences, Macquarie University, NSW 2109, Australia 
}

   \date{Received XXX; accepted YYY}

\abstract
{
The long GRB\,180728A, at a redshift of $z=0.1171$, stands out due to its high isotropic energy of $E_\mathrm{\gamma,iso}\approx2.5\times10^{51}$ erg, 
in contrast with most events at redshift $z<0.2$, but comparable to the bulk of luminous bursts more common at higher redshift.
}
{
We analyze the properties of GRB\,180728A's prompt emission, afterglow, and associated supernova (SN 2018fip), comparing them with other GRB-SN events.
}
{
This study employs a dense photometric and spectroscopic follow-up of the afterglow and the SN up to 80 days after the burst, supported by image subtraction to remove the presence of a nearby bright star, and modelling of both the afterglow and the supernova.
}
{
GRB 180728A lies on the $E_\mathrm{p,i}$--$E_\mathrm{\gamma,iso}$ plane occupied by classical collapsar events, and the prompt emission is one of the most energetic at $z<0.2$ after GRB 030329 and GRB 221009A.
The afterglow of GRB\,180728A is less luminous than that of most long GRBs, showing a shallow early phase that steepens around 5 hours (0.2 days). 
The GRB exploded in an irregular, low-mass, blue, star-forming galaxy, typical of low-$z$ collapsar events. Because of the relatively faint afterglow, the light curve bump of SN\,2018fip dominates the optical emission already after $\sim3$~days and is one of the best sampled to date. 
The strong suppression below $\sim4000$~\AA{} and a largely featureless continuum in the early 6--9 days spectra favor
aspherical two-component ejecta with a high-velocity collimated component ($>20{,}000$ km s$^{-1}$), dominant early-on, and a more massive, low-velocity component, which dominates at much later epochs.
}
{
Our findings indicate that asymmetries need to be considered in order to better understand GRB-SNe.
In any case, SN\,2018fip shares many characteristics with typical GRB-SNe. Its kinetic energy is below the common range of $10^{52}$--$10^{53}$~erg 
and does not correlate with the high energy of the GRB, highlighting the complexity and diversity of the GRB-SN energy budget partition. 
}


\maketitle

\section{Introduction}

Gamma-ray bursts (GRBs) are the most energetic explosions in the Universe.
They have at least two different progenitors: either the collapse of fast-rotating very massive stars \citep[collapsar model, e.g.,][]{WoosleyBloom2006a} which leads to broad-lined Type Ic core-collapse supernovae (SNe) \citep{Cano2017a}, or the merger of compact objects including at least one neutron star, which can be observed as kilonovae \citep[KNe; e.g.,][]{Abbott2017a,Abbott2017b}. 
Although some exceptions have been found \citep[e.g.,][]{Ahumada2021a,Rossi2022a, Levan2023a}, collapsar events are usually responsible for long-duration GRBs (LGRBs), which have a duration longer than $\sim2$ s \citep[see, e.g.][]{Bromberg2012a} and a soft energy spectrum \citep{Mazets1981a,Kouveliotou1993a}.

The association of LGRBs with massive stars has been established
since the association of the nearby and low-energy GRB 980425 with
the well studied SN 1998bw, the prototypical GRB-SN \citep[e.g.,][]{Galama1998a,Woosley1999a,Nakamura2001a,Sollerman2002a,WoosleyBloom2006a,Clocchiatti2011AJ,Modjaz2016a}.
All known GRB-SNe are Type Ic with broad line features (SNe Ic-BL), explosions of highly stripped stars that lack signatures of hydrogen and helium in their spectra \citep[e.g.,][]{WoosleyBloom2006a,Cano2017a}. 
Most GRB-SNe share similar luminosity\footnote{But note that Type Ic-BL SNe without confirmed GRB are in average fainter than GRB-SNe and SN 1998bw in particular \citep[e.g.,][]{Taddia2019a,Srinivasaragavan2024b}.}, energy release, ejecta and nickel masses \citep[e.g.,][]{Cano2017a,Izzo2019a,Klose2019a,Melandri2022a}, regardless of the energy of the GRB, whether they are a low-energy\footnote{Note that other works classify GRBs as low or high gamma-ray luminosity events. However, here we prefer to consider just the energy to better compare with the total energy-budget available to the GRB-SN event.} GRB such as GRB 060218 \citep[e.g.][]{Pian2006Nature,Mazzali2006Nature}, or a highly energetic
GRB such as GRB 130427A \citep[e.g.,][]{Xu2013a,Melandri2014a}. 
The only rare exceptions are the superluminous SN 2011kl associated with the ultra-long GRB 111209A \citep{Kann2019AA}, and the putative most luminous GRB-SN associated with the otherwise unexceptional GRB 140506A\footnote{Though it is not spectroscopically confirmed.} \citep{Kann2024a}.

In contrast to the GRB-SN properties, LGRBs have a wide range of emitted isotropic energy \citep[$E_\mathrm{\gamma,iso}$ of $10^{48}$--$10^{55}$ erg; e.g.,][]{Minaev2020a,Tsvetkova2021a}. 
Due to selection effects \citep[e.g.,]{Minaev2020a},
most events within $z<0.2$ are low-energy GRBs, with an isotropic energy of $10^{48}$--$10^{50}$ erg, while high-redshift events have large gamma-ray energies, $10^{51}$--$10^{55}$ erg \citep[e.g.,][]{Minaev2020a,Tsvetkova2021a}. 
Among the dozen LGRBs at $z<0.2$, only two have $E_\mathrm{\gamma,iso}>10^{50}$ erg, 
GRB 030329 at $z=0.167$ \citep{Hjorth2003b,Matheson2003ApJ,Stanek2003ApJ}, and the exceptional Brightest Of All Time (BOAT) GRB 221009A \citep{Burns2023a} at $z=0.151$ \citep{Malesani2025a} with its $E_\mathrm{\gamma,iso}\approx10^{55}$ erg \citep{Frederiks2023a}. 

Of the thousands of known LGRBs, 
we have identified an associated SN
in only $\sim$50 cases through late-time bumps in their optical afterglow light curves, since such signals become progressively too faint to detect at high redshifts.
Moreover, we have accurate spectroscopic observations of only about 30 GRB-SNe, because we can only study the low-redshift events in detail \citep[e.g.,][]{Galama1998a,Hjorth2003b,Izzo2019a,Cano2017b,Ashall2019a, Melandri2019a,Kann2019AA,Melandri2022a}.
SNe associated with events at redshifts beyond $z=0.2$ are usually too faint for spectroscopic follow-up from the ground except for the brightest events at $z\gtrsim$0.3, such as GRB 130427A - SN 2013cq at $z=$0.3399 \citep[e.g.,]{Xu2013a,Melandri2014a} , or more recently GRB 230812B - SN 2023pel $z=$0.360 \citep[][]{Srinivasaragavan2024a,RomanAguilar2025a}, and even their photometric follow-up requires a substantial observational effort \citep[e.g.,][]{Klose2019a}.

GRB 030329 is the only nearby high-energy event whose afterglow did not outshine its associated SN, unlike GRB 221009A
\citep{Shrestha2023a,Levan2023a,Blanchard2024a,Srinivasaragavan2023a}, allowing detailed study from its rise. With just this one case, the connection between GRB energy and SN properties remains uncertain.
Within the collapsar scenario, the observed burst duration is just the difference between the engine operating time and the jet breakout time \citep[e.g.,][]{Bromberg2012a}. Numerical simulations showed that
the longest-lasting engines result in the most successful gamma-ray bursts (the result is also a function of the viewing angle), those in which the jet breaks out of the star’s surface \citep{Lazzati2012a,Lazzati2013a}. 
Most of the energy produced powers the SN ejecta, while only a small fraction is sufficient for the jet to penetrate the stellar envelope and produce the GRB and the afterglow \citep[e.g.,][]{Mazzali2014a,Ashall2019a}. 
However, the simple assumption of a spherical explosion is not the best model. In particular, \citep{Ashall2019a} found that the emission of SN 2016jca (GRB 161219B) is a highly aspherical explosion viewed close to the on-axis jet, due to the very early SN spectroscopic identification. \citep{Izzo2019a} reached similar conclusions in the case of SN 2017iuk (GRB 171205A). 

In this paper, we present a comprehensive analysis of the optical and near-infrared counterpart of GRB 180728A and of the associated supernova. This event has a gamma-ray energy of $\approx 2 \times 10^{51}$ erg \citep{Frederiks2018GCNa},
firmly placing it in the high-energy population of GRBs.
In \citep{Rossi2018GCN180728}, we reported the detection in our first afterglow spectra of absorption features at a common redshift of $z=0.117$ (refined to 0.1171 in section \ref{subsec:spectra}), which we consider to be the redshift of GRB 180728A. 
In \citet{Izzo2018GCNa} and \citet{Selsing2018GCNa}
we have reported the spectroscopic discovery of the emerging SNa, named SN 2018fip in the Transient Name Server.\footnote{\href{https://wis-tns.weizmann.ac.il/object/2018fip}{https://wis-tns.weizmann.ac.il/object/2018fip}} 

In Sections \ref{sec:Observations} and \ref{sec:Data_reduction}, we present the available multi-band data; in Section \ref{sec:Data_analysis}, we present the results of our analysis of the afterglow, host galaxy, and SN spectra; in Section \ref{sec:Discussion}, we discuss our analysis; in Section \ref{sec:summary_and_conclusions}, we present our conclusions.
Throughout this work, we adopt the notation according to which the flux density of a counterpart is described as $F_\nu (t) \propto t^{-\alpha} \nu^{-\beta}$ 
and we use a $\Lambda$CDM world model with $\Omega_M = 0.308$, $\Omega_{\Lambda} = 0.692$, and $H_0 = 67.8$ km s$^{-1}$ Mpc$^{-1}$ \citep{Planck2016a}. 
The redshift of $z=0.117$ thus corresponds to a luminosity distance of $562$ Mpc, and at this redshift 1 arcsec corresponds to 2.2 kpc.
Significant Galactic extinction affects the field 
\citep[A$_{\rm V} = 0.74$ mag;][]{SchlaflyFinkbeiner2011a}.
We report magnitudes in the AB system. 
We report times in the observer frame, unless otherwise specified.

\begin{figure}[tp]
\begin{center}
\includegraphics[height=0.45\textwidth,angle=0]{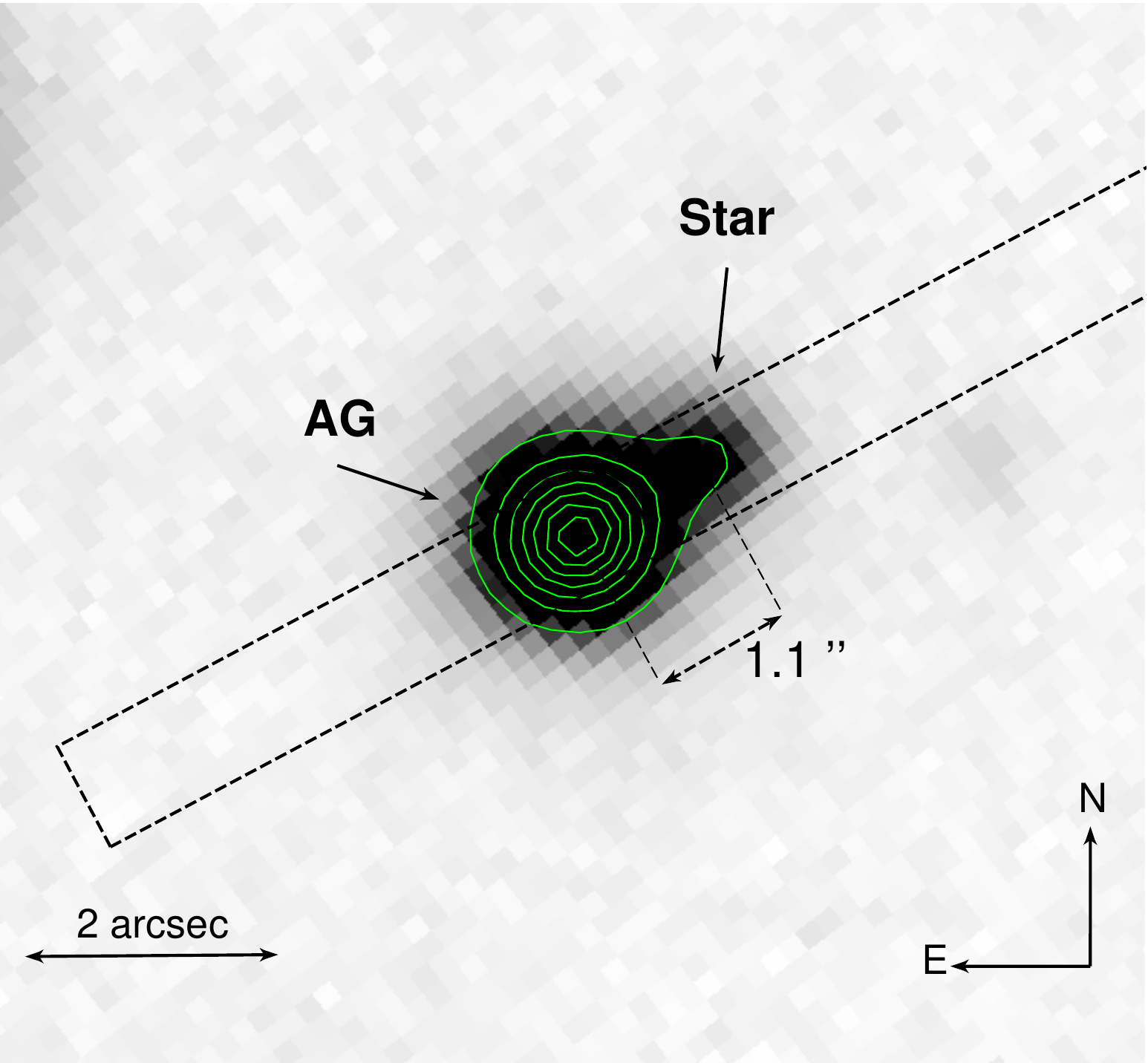}
\caption{
$r$-band image of the optical afterglow of GRB 180728A obtained with the X-shooter acquisition camera on 2018-07-28, 0.23 days after the burst trigger. We highlight the position of the afterglow (AG) and of the nearby star, together with their projected distance.
The rectangle shows the X-shooter slit with a position angle of 118.1 deg, which we selected to cover both the AG and the nearby star starting with the 4th observation at 6.26 days. The green contour lines indicate equal count levels.}
\label{fig:field}
\end{center}
\end{figure}

\section{Observations}\label{sec:Observations}

\subsection{The burst}\label{sec:burst}

The Burst Alert Telescope \citep[BAT;][]{Barthelmy2005a} on board the Neil Gehrels \textit{Swift} Observatory satellite \citep{Gehrels2004a} detected the bright LGRB 180728A at $t_0=$17:29:00 UT on the 28th of July 2018 \citep{Starling2018GCNa}. 
The prompt emission shows a faint precursor lasting about 3 s followed by a single but brighter pulse. This began at $t_0+11$~s, peaked at $t_0+13$~s, and faded to background at $t_0+40$~s. The total duration of the burst measured by BAT in the $15$--$350$~keV energy range is $T90=8.68\pm0.30$~s \citep{Markwardt2018GCNa}.
Konus-Wind (KW; \citealt{Frederiks2018GCNa}) and
 {\it Astrosat} CZTI \citep{Sharma2018GCNa} also detected the burst, 
 while the Gamma-Ray Burst Monitor (GBM) onboard the {\it Fermi} satellite \citep{Veres2018GCNa} detected its precursor.

\subsection{Afterglow photometry}\label{sec:ag}

The MASTER Global Robotic Net \citep{Lipunov2010a}  
pointed at GRB 180728A with MASTER-SAAO, located at the South African Astronomical Observatory starting 22~s after notice time (38~s after trigger time) on 2018-07-28 17:29:38 UT \citep{Lipunov2018GCNa,Lipunov2018GCNb}. 
In the first 10~s exposure the MASTER auto-detection system discovered the optical afterglow in both polarization filters (see Sect.~\ref{sec:dataphot} where we provide a refined localization). 
MASTER followed-up the afterglow every night until 2018-08-18 with both MASTER-OAFA (located at the Observatorio Astronomico Felix Aguilar) and SAAO \citep{Lipunov2018GCNc}.

We conducted additional early optical and near-infrared (NIR) observations using the 0.6-meter robotic REM (Rapid Eye Mount) telescope \citep{Zerbi2001a,Covino2004a}, located at the European Southern Observatory (ESO) in La Silla, Chile. Observations began on 2018-07-28, about 0.18 days after the burst event, and lasted for several hours. 
We performed further ground-based observations using the multichannel imager GROND \citep{Greiner2008a}
mounted on the MPG 2.2m telescope on La Silla, Chile.
GROND started observing on 2018-08-01, 3.3 days after the GRB trigger.
Starting 5.6 hours after the trigger, we have also used the acquisition camera of the X-shooter instrument. 

{\it Swift} did not slew immediately due to an Earth limb constraint \citep{Starling2018GCNa}.
The X-Ray Telescope \citep[XRT;][]{Burrows2005a} and the UltraViolet and Optical Telescope \citep[UVOT;][]{Roming2005a} aboard \textit{Swift} began observing GRB 180728A 1730.8 s (0.02 days) after the BAT trigger \citep{Perri2018GCNa}.
{\it Swift}/XRT found an unknown X-ray source
with an uncertainty of 5.8 arcseconds (radius, 90\% containment).
A UVOT-enhanced position gave the localization to within 1\farcs5 radius ($90\%$ containment) 
and was consistent with the optical counterpart \citep{Osborne2018GCNa}. XRT observations continued for more than 2 months after the GRB, when the source became too faint to be detected. 
${\it Swift}$ UVOT began settled observations of the field 1740\,s after the BAT trigger, and initial results were reported in \citet{Laporte2018aGCN}, where they report the detection of a source consistent with the enhanced XRT position and the optical transient (OT) discovered with MASTER. 

\subsection{Spectroscopic follow-up} 

We obtained UV to NIR spectroscopic observations of the afterglow and SN with the X-shooter instrument \citep{Vernet2011a} mounted on the 
Very Large Telescope (VLT) on Paranal (ESO, Chile). The spectra cover a wavelength range from 3300-22500 \AA. 
We obtained the first three spectra under the ESO program 0101.D-0648 (PI: N. Tanvir), and the following spectra under the DDT ESO program 2101.D-5044 (PI: A. Rossi).

Starting with the fourth epoch, we placed the slit at a position angle of $118.1^\circ$, chosen to cover both the AG and the nearby star, as shown in Fig. \ref{fig:field}. Note that by chance the slit covered both objects also during the first observation. Observations were obtained by nodding along the 
slit with an offset of $5''$ between exposures in a standard ABBA sequence. We used slit widths  of 
$1\farcs0$, $0\farcs9$ and $0\farcs9$ for the UVB, VIS and NIR spectrograph arms, respectively, resulting in resolving powers of $R = \lambda / \Delta \lambda \approx 4400, 7400$ and $5400$.
Our spectroscopic campaign is summarized in Table \ref{tab:spec}. 
\citep{Selsing2018GCNa} reported that significant features in the spectrum develop and become more prominent 23 days after the GRB (21 days in the rest frame). 
These features resemble those of SNe Type-Ic at maximum light (see Sect.~\ref{sec:bolo}).

\begin{table}
\begin{center}
\caption{List of the spectroscopic observations obtained with X-shooter.\label{tab:spec}}
\begin{tabular}{cccccc} 
\toprule
 MJD    & Time$^a$   & $t_{exp}$  & Seeing & Airmass & PA$^b$ \\
 $[$day$]$      & $[$day$]$ & $[$s$]$      & $[^{\prime\prime}]$   &         &   $[$deg$]$    \\
\midrule
58327.97 &0.24   & 1x600  & 0.84  &  1.24 &  52.83 \\
58328.22 &0.50   & 2x600  & 0.93  &  1.70 & -79.76 \\
58329.21 &1.48   & 2x600  & 0.99  &  1.67 & -75.59 \\
58333.99 &6.26   & 2x600  & 0.75  &  1.16 & -118.1 \\
58337.05 &9.32   & 4x600  & 0.85  &  1.16 & -118.1 \\
58340.02 &12.29  & 4x600  & 0.78  &  1.15 & -118.1 \\
58346.10 &18.37  & 4x600  & 0.96  &  1.23 & -118.1 \\
58351.09 &23.36  & 4x600  & 0.89  &  1.24 & -118.1 \\
58369.05 &41.32  & 4x600  & 0.77  &  1.30 & -118.1 \\
58399.00 &71.27  & 4x600  & 1.02  &  1.58 & -118.1  \\
58403.02 &75.29  & 4x600  & 1.18  &  1.64 & -118.1  \\
58405.02 &77.29  & 4x600  & 0.59  &  1.97 & -118.1  \\
58406.01 &78.28  & 4x1200 & 0.84  &  1.77 & -118.1  \\
\bottomrule
\end{tabular}
\end{center}
\begin{tablenotes}
\footnotesize
\item $^a$ {Times are the midtime after burst trigger.} 
\item $^b$ {PA indicates the position angle from N to E.}
\end{tablenotes}
\end{table}


\section{Data reduction}\label{sec:Data_reduction}

\subsection{Imaging} \label{sec:dataphot}

We extracted ${\it Swift}$ UVOT source counts using a region of 3\arcsec\ radius. In order to be consistent with the UVOT calibration, we then corrected these count rates to 5\arcsec\ using the curve of growth contained in the calibration files \citep{Poole2008a}. 
Within this extraction region there is contamination from a relatively bright nearby star in the UVOT $white$, $b$ and $v$ filters (see Fig.~\ref{fig:field}), 
which we subtracted using late-time imaging. The star was too faint in the UV and not a significant source of contamination at those wavelengths.
We obtained the count rates from the images using the ${\it Swift}$ tool {\sc uvotsource} and converted to magnitudes using the UVOT photometric zero points.
Contamination-corrected fluxes can be found in Table \ref{tab:photalt}. 

MASTER observed the OT in both polarization filters ($Pola1$ and $Pola2$ in Tab. \ref{tab:photalt}) and in Clear $CR$ band (best described by GAIA $g$-ﬁlter). 
Unfortunately, we could not measure any polarization.
The list of reference stars is in Table \ref{tab:master-stars}. 
After astrometric calibration and combine of the images, we performed standard aperture photometry as described in \citep{Lipunov2019a}. In the attempt to detect the SN component, deep observations in the Clear-band were obtained after the first night. These frames were stacked and an upper limit was calculated for each of the resulting total frames.

REM data reduction was performed using standard procedures, including image alignment, stacking, and sky subtraction. 
The images were automatically reduced using the jitter script of the \texttt{eclipse} package \citep{Devillard1997a} which  aligns and stacks the images to obtain one average image for each sequence. A combination of IRAF \citep{Tody1993} and SExtractor packages \citep{bertin2010sextractor} was used to perform aperture photometry.
Given the relatively low signal-to-noise ratio of the individual images, we combined multiple frames, obtaining a significant detection in the SDSS $g^\prime r^\prime i^\prime z^\prime$ filters at $0.25$ days.

X-shooter $g^\prime r^\prime z^\prime$ and GROND $g^\prime r^\prime i^\prime z^\prime JHK_s$ images were reduced in a standard manner using PyRAF/IRAF \citep{Tody1993}. 
In particular, for GROND, a dedicated pipeline was used as described in \citet{Kruhler2008a}.
Astrometry was calibrated against field stars in the GAIA DR2 catalog \citep{Gaia2018a}, obtaining an astrometric precision of 0\farcs17.
The coordinates of the transient are 
RA,DEC(J2000)= $+16^{\rm h}54^{\rm m}15\fs 48$, $-54^\circ 02'40\farcs3$.
In GROND and X-shooter images the nearby star makes it difficult to perform photometry, especially when the OT is fainter than the star and the seeing is comparable or larger than the projected offset (Fig.\ref{fig:field}). 
To remove the contamination of this star, we used image-subtraction using deep X-shooter and GROND reference images obtained more than 220 days after the GRB under clear sky conditions with seeing $\sim0\farcs9$ and $\sim0\farcs8$ for the X-shooter and GROND, respectively.
Before applying image subtraction, the input and reference images were aligned using the \texttt{WCSREMAP} package \citep{Mink1997a}. Image subtraction was then performed using a routine based on
\texttt{HOTPANTS}\footnote{\url{https://github.com/acbecker/hotpants}} \citep{Becker2015a}. 
To calibrate the $g'r'i'z'$ photometry we used secondary standards (Tab~\ref{tab:stdgrond}) 
in the field observed with GROND 16.3 days after trigger, and calibrated with the SDSS field STD17\footnote{Centred at 
RA,DEC(J2000)= $+17^{\rm h}00^{\rm m}22\fs 5$, $-11^\circ 17'25$.}
observed immediately after and under photometric conditions. 
Calibration of the field in $JHK_s$ was performed using 2MASS stars \citep{Skrutskie2006a}.

Finally, we have corrected all data for Galactic extinction using the extinction curve derived by \citet{Cardelli1989},   
$E(B-V)=0.238$ mag from the dust maps of \citet{SchlaflyFinkbeiner2011a},
and an optical total-to-selective extinction ratio $R_V=3.1$. 
In table 
\ref{tab:photalt} we report our full data set, before correction for Galactic extinction.

\subsection{Spectroscopy of the optical transient}
\label{subsec:spectra}

We reduced the spectra following the procedure described in \citet{Selsing2019a}, 
which includes a cosmic-ray removal algorithm \citep{vanDokkum2001a} applied to the raw spectra, after which each individual exposure was reduced with the version {\texttt v. 3.5.0} of the ESO X-shooter pipeline \citep{Modigliani2010a}. The pipeline produces a flat-fielded, rectified, and wavelength-calibrated 2D spectrum for every frame in the UVB, VIS, and NIR arms. We then combined the different frames using custom-made post-processing scripts\footnote{\url{https://github.com/jselsing/xsh-postproc}}. 
Background light from a nearby bright foreground star complicated the extraction of the 1D spectrum.
We used an extraction region from $-3$ to $+3$ pixels (equivalent to 0.96'') to limit the contamination from the stellar source. 
The final extracted 1D spectra were then corrected for slit loss and Galactic extinction along the line-of-sight of the burst using the dust maps of \citet{SchlaflyFinkbeiner2011a}. 
Wavelengths are reported in vacuum and in the barycentric frame of the Solar System.
After analyzing the final reduction of the first spectra, we have refined the redshift to be $z=0.1171\pm0.0001$, determined from the detection of absorption features due to
Mg\,$\rm II \, \lambda\lambda$2796,2803, Mg\,$\rm I\,\lambda$2853, and Ca\,$\rm II\,\lambda\lambda$3934,3969,
as first reported in \citet{Rossi2018GCN180728}.
The absolute-flux calibrated and afterglow subtracted spectra of SN 2018fip are available on the Weizmann Interactive Supernova Data Repository\footnote{\url{https://www.wiserep.org}.} (WISeREP). 


\begin{figure}[t!]
\begin{center}
\includegraphics[width=0.5\textwidth,angle=0]{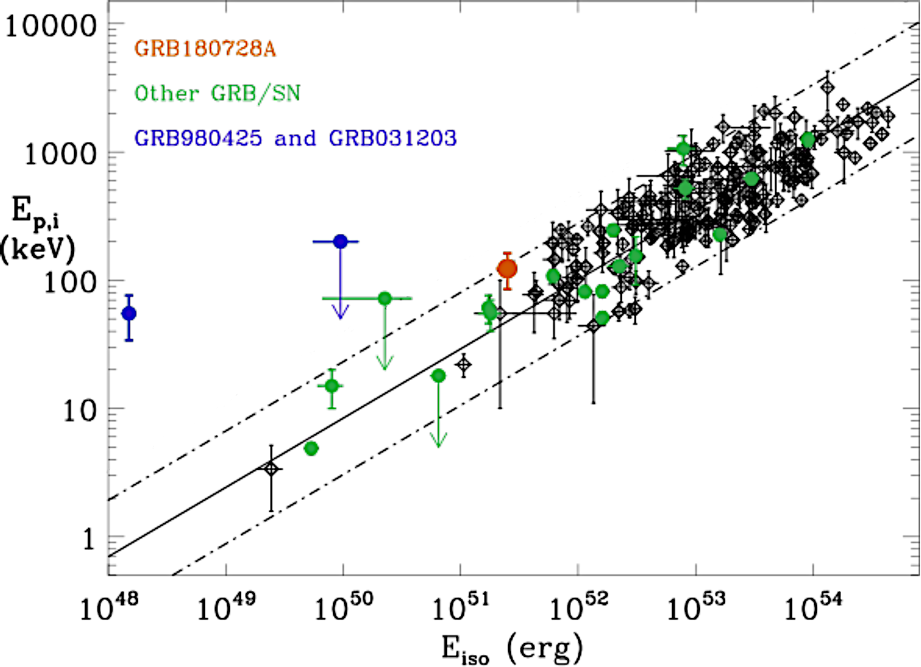}
\caption{GRB 180728A (red) in the $E_\mathrm{p,i}-E_\mathrm{\gamma,iso}$ plane. GRBs with an associated SN are highlighted in green, outliers in blue. Dot-dashed lines indicate the $\pm2\sigma$ region. GRB data are from \citet{Amati2019a}. 
}
\label{fig:amati}
\end{center}
\end{figure}

\section{Data analysis}\label{sec:Data_analysis}
\subsection{Prompt emission phase} \label{sec:grbene}

The prompt emission of GRB 180728A was measured and characterized by the three main GRB detectors currently in operation: 
Swift/BAT 
\citep[][]{Markwardt2018GCNa}, Fermi/GBM \citep[][]{Veres2018GCNa}, and Konus-WIND \citep[][]{Frederiks2018GCNa}.
The light curve of this event consists of a weak and soft ``precursor'' followed by a bright and harder pulse \citep[see also][]{Hu2021a}, 
making it one of those cases in which the estimates of the observer-frame spectral peak energy $E_\mathrm{p}$ and, to a lesser extent, of the total radiated energy $E_\mathrm{\gamma,iso}$ depend significantly on the combination of the energy band and detector sensitivity, as well as on the exposure time over which the spectrum and fluence were measured.

In order to check the consistency of this event with the $E_\mathrm{p,i}$-$E_\mathrm{\gamma,iso}$ correlation\footnote{$E_{\rm p,i}=E_{\rm p}(1+z)$ is the  rest-frame photon  energy at which the $\nu$F$_\nu$ spectrum peaks, and $E_\mathrm{\gamma,iso}$ is the isotropic-equivalent radiated energy as measured in a ``bolometric'' band, usually 1 keV -- 10 MeV in the rest frame.} of LGRBs \citep{Amati2002a,Amati2006a}, we adopted the conservative approach of taking into account the fluence and $E_\mathrm{p}$ estimates for all three instruments.
In the case of BAT we performed our own analysis, following the standard data reduction and analysis recipes \footnote{BAT Analysis Threads on the \url{nasa.gov} website}. 
Based on the above and assuming the standard flat $\Lambda$CDM cosmology, 
we derived an isotropic equivalent radiated energy $E_\mathrm{\gamma,iso}$ in the 1--10000 keV cosmological rest-frame of $(2.5\pm0.5)\times10^{51}$ erg and a rest-frame, intrinsic spectral peak energy E$_\mathrm{p,i}$ of $123\pm28$ keV. As shown in Figure \ref{fig:amati}, these values make GRB 180728A well consistent with the $E_\mathrm{p,i}$-$E_\mathrm{\gamma,iso}$ correlation \citep[e.g.,][]{Amati2002a}. 
With the values above, GRB 180728A had an isotropic gamma-ray equivalent luminosity of $\log L_\mathrm{iso}$ [erg s$^{-1}$] = 50.4
and falls in the category of high-luminosity GRBs \citep{Hjorth2013a, Cano2017a}.

\begin{figure}[!tp]
\begin{center}
\includegraphics[width=0.49\textwidth,angle=0]{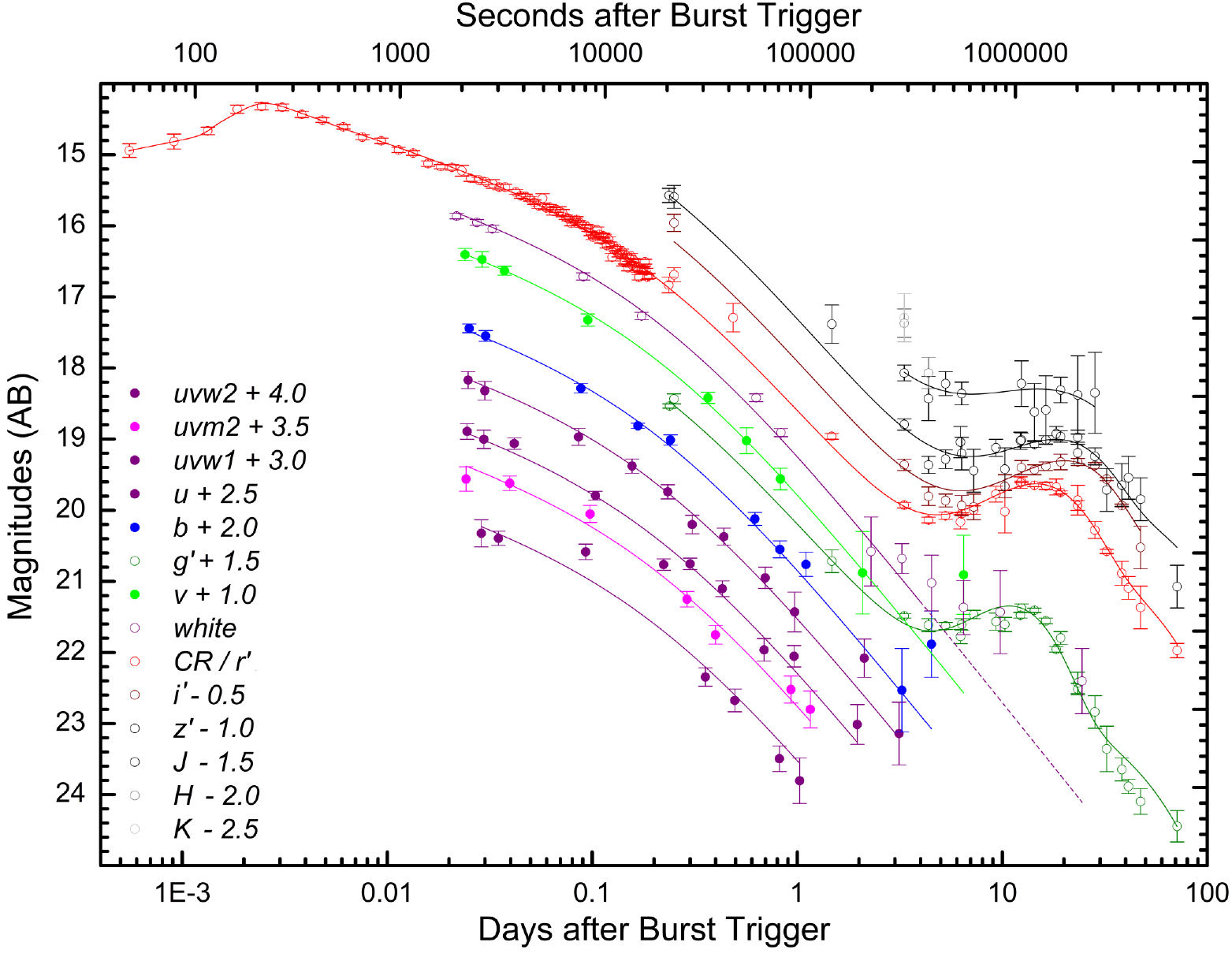}
\caption{Fit to the light curve and supernova of GRB 180728A. Magnitudes are in the AB system and corrected for Galactic foreground extinction, and time is in observer-frame. For reasons of clarity, light curves in each filter are offset by the magnitude values noted in the legend. Starting at $\approx300$ s, the light curve in all bands except for $H$ and $K$ is fitted with a broken power-law plus individual SN component. The dashed line for the $white$ filter is an extrapolation because the points at $>4$ days were not fitted. See text for details.}
\label{fig:SNfit}
\end{center}
\end{figure}

\begin{table}[t]
\centering
\caption{Afterglow and Supernova Fit Results}
\begin{tabular}{l c c }
\toprule
Parameter & $n^a=1$ \\
\midrule
$\chi^2/$d.o.f.	&	2.214			 		\\
$\alpha_1$	& $	0.367	\pm	0.011	$ \\
$\alpha_2$	& $	1.451	\pm	0.020	$ \\
$t_b$ (day)	& $	0.215	\pm	0.016	$ \\
\midrule
$k_{g^\prime}$	& $	0.884	\pm	0.019	$ \\
$s_{g^\prime}$	& $	0.767	\pm	0.011	$ \\
$k_{r^\prime}$	& $	0.775	\pm	0.007	$ \\
$s_{r^\prime}$	& $	0.854	\pm	0.008	$ \\
$k_{i^\prime}$	& $	0.779	\pm	0.019	$ \\
$s_{i^\prime}$	& $	1.034	\pm	0.018	$ \\
$k_{z^\prime}$	& $	0.772	\pm	0.022	$ \\
$s_{z^\prime}$	& $	0.976	\pm	0.047	$ \\
$k_{J}$	& $	1.331	\pm	0.171	$ \\
$s_{J}$	& $	0.956	\pm	0.286	$ \\
\bottomrule
\end{tabular}
\begin{tablenotes}\footnotesize
\item $^a$ $n$ is the smootheness parameter of the smoothly broken power-law after the peak of the light curve. See text for details.
\end{tablenotes}
\label{tab:lcfitpara}
\end{table}

\subsection{Light curve analysis -- the afterglow and supernova}
\label{sect:lcana}

The very early MASTER light curve shows a shallow, then steeper, rise to peak.
To model this peak, we excluded the first two data points, 
where residual prompt emission might still be present.
Therefore, we modelled data up to 0.05 days after the trigger with a rise breaking to a decay using a smoothly broken power-law \citep{Beuermann1999a}: 
$F = (F_1^{n}+ F_2^{n})^{-1/n}$,
where $F_\textrm{x}=f_\textrm{b}(t/t_\textrm{b})^{-\alpha_x}$\footnote{Note that we define $F_\nu (t) \propto t^{-\alpha}$, thus a negative temporal index indicates a rising light curve.},
$f_\textrm{b}$ being the flux density at break time $t_\textrm{b}$, $n$ the break smoothness parameter, and the subscripts $1,2$ indicate pre- and post-break, respectively. In particular, using the subscript r,d for the rise and decay around the early peak, we
find $\alpha_\mathrm{r}=-0.66\pm0.16$, $\alpha_\mathrm{d}=0.431\pm0.011$, and $t_\mathrm{b,peak}=0.0024\pm0.0002$ days ($211\pm16$ s). 

Data following the peak can also be fit with a smoothly broken power-law, in addition to an SN component (the underlying host-galaxy component has been subtracted). 
As previously reported in \citet{Izzo2018GCNa}, there is clear evidence of a $\sim0.5$ mag rebrightening in the X-shooter $r$-band between 6 and 13 days,
thus indicating an emerging SN component in the optical light curve. 
Therefore, we fit the afterglow of GRB 180728A as well as SN 2018fip in the $k,s$ context \citep{Zeh2004ApJ}. 
Here, we assume that the SN light curve evolves like SN 1998bw associated with GRB 980425 \citep{Galama1998a,Patat2001a,Clocchiatti2011AJ}, modified by the luminosity factor $k$ and the stretch factor $s$.
A value $k=1$ implies that the
GRB-SN is just as luminous as SN 1998bw in the specific rest-frame band corresponding to the
observer-frame band in which the measurements were taken. Furthermore,
without changing its fundamental shape, the light curve's temporal
evolution can be compressed ($s<1$) or stretched ($s>1$) with the
stretch factor, $s$. 
This is a powerful analysis method, as GRB-SNe are found to generally agree well with the light curve shape of SN 1998bw \citep{Ferrero2006AA,Klose2019a,Kann2024a}.

We performed a joint fit of all bands except the GROND $H$ and $K$ bands, which have so few detections that they do not contribute to the fit. In this fit, the parameters $\alpha_1$, $\alpha_2$, $t_b$ and $n$ are shared between all bands. Similarly, we assume no host-galaxy contribution in any band. We do not find any systematic offsets between Sloan $g^\prime r^\prime z^\prime $ from X-shooter, GROND $g^\prime r^\prime z^\prime $ data and MASTER $CR$ data, therefore we combine similar filters into single light curves. 
Note that only GROND systematically observed in $i^\prime$ and the NIR filters, except for one REM epoch with $g^\prime r^\prime i^\prime z^\prime $ photometry at $\sim0.25$ days.
As GROND data dominates especially during the SN phase, we will henceforth use these filters when filter-specific parameters are needed. For all bands where a SN contribution is detected (GROND $g^\prime r^\prime i^\prime z^\prime J$), we derive SN 1998bw model light curves in these filters at the redshift of GRB 180728A following the method detailed in \citet{Zeh2004ApJ} and \citet{Klose2019a}. 
The luminosity factor $k$ and the stretch factor $s$ are left free to vary individually for each band. The derived $k_{g'...J}$ parameters represent the exact SN 1998bw light curve for each filter.
As the data density is high, we initially decided to leave the break smoothness $n$ free to vary. However, this resulted in a degenerate fit which did not converge after over a thousand Levenberg-Marquardt least-squares iterations.
Because the light curve break clearly shows a very smooth rollover, we fixed $n=1$.
The results are given in Table \ref{tab:lcfitpara}.\footnote{As the determined $\alpha_{1,2}$ values are those for $t\rightarrow0$ and  $t\rightarrow+\infty$, respectively (and thus difficult to visualise the difference), the slope results differ depending on the choice of $n$.} 
The $n=1$ fit is also more consistent with the expected values for a typical GRB geometry \citep{vanEerten2013a, Lamb2021a}.
The break is achromatic, a characteristic feature of jet breaks \citep[e.g.,][]{Rhoads1997a,Rhoads1999a}, which is clear in the optical bands (Fig.~\ref{fig:SNfit}). We can therefore exclude an origin related to the passage of a spectral break \citep[e.g.,][]{Sari1998a}. 
Alternative scenarios such as reverse shocks or a sudden energy injection would instead produce a bump or a steep-to-shallow transition in the light curve \citep[e.g.,][]{SariMeszaros2000a,ZhangMeszaros2002a,Granot2003a,NakarGranot2007a}. 
We therefore conclude that this feature is most likely the jet break of the afterglow. The pre- and post-break decay indices are shallower than expected for a classical jet break, and may be explained by a non-constant environment (e.g., wind) with the possible presence of continuous (rather than sudden) energy injection \citep[e.g.,][]{Zhang2006c,Racusin2009a} and/or by a slightly off-axis structured jet \citep[e.g.][]{Ryan2020a}. An advanced numerical modelling goes beyond the goals of this work. 

Given the very low line-of-sight extinction in the host galaxy of GRB 180728A (see Sect.~\ref{sect:SED}), the $k$ values derived from the fit require no further correction. 
We find SN 2018fip is generally somewhat fainter than SN 1998bw, 80\% -- 90\% of its luminosity, except for the $J$ band, where it is somewhat brighter (but we caution the error here is larger, and SN 1998bw itself has very sparse and late $J$-band data \citealt{Patat2001a,Sollerman2002a}). It also evolves a bit faster than SN 1998bw, at a stretch of 80\% -- 100\%. This makes SN 2018fip a very typical GRB-SN.
We also note that this SN joins the small group of GRB-SN with NIR detections, which includes SN2010bh, SN2011kl, SN2013dx, and SN1998bw.

The fitted model allows us to disentangle the afterglow and SN contributions from the data 
producing high-quality ``pure'' AG and SN-only data.\footnote{Analog to the treatment of the afterglow of GRB 111209A and SN 2011kl \citep{Greiner2015Nature,Kann2018AA,Kann2019AA}}

\begin{figure}[!tp]
\begin{center}
\includegraphics[width=0.5\textwidth,angle=0]{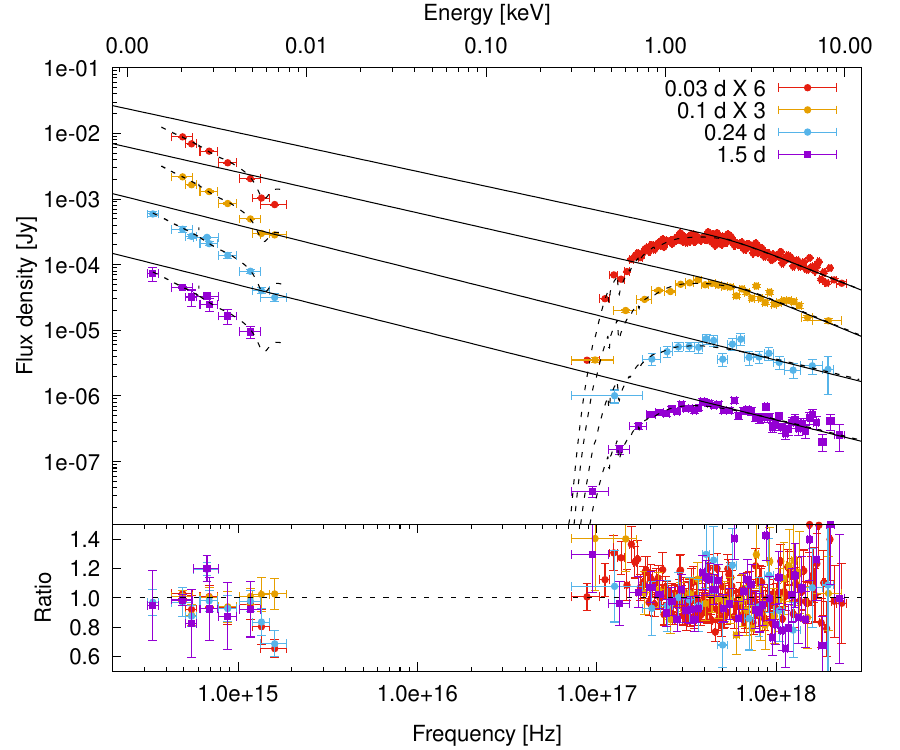}
\caption{X-ray to optical SED at 0.03, 0.1, 0.24 and 1.5 days.
The last 2 epochs are well modelled by a joint-fit with a single power-law with spectral slope $\beta\sim0.7$. In the first 2 epochs the optical data has a slightly shallower spectral slope $\beta_{opt}\sim0.6$, but not the X-ray data, and needs a break that moves from $2.1$ to $2.6$ keV (with $\beta_{X}=\beta_{opt}+0.5$). 
The solid line is the unextinguished/unabsorbed model. The dashed line represents the extinguished/absorbed model, dominated by the Galactic foreground extinction in the optical/NIR bands and MW absorption in the X-rays.} 
\label{fig:sedbox}
\end{center}
\end{figure}


\begin{figure}[!thp]
\begin{center}
\includegraphics[width=0.5\textwidth,angle=0]{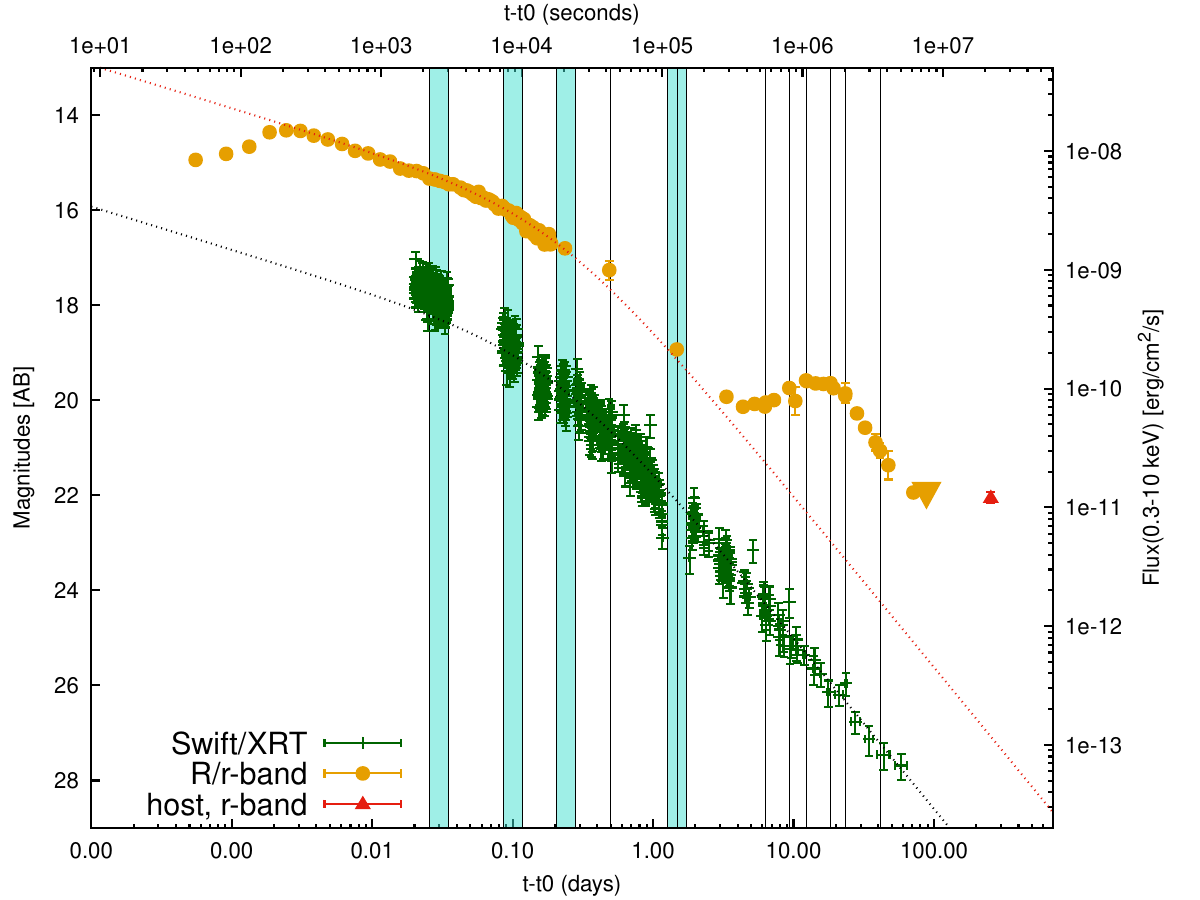}
\caption{The {\it Swift}/XRT and optical light curves (MASTER CR band, and X-shooter, GROND $r^\prime$). Note that the optical data of the transient was obtained via image subtraction after 0.2 days, and therefore the contribution of the host is also subtracted. 
We have highlighted in cyan the epochs used in the optical-to-XRT SED fitting. Vertical lines highlight epochs with spectroscopic coverage. The dotted lines are the best fit of the optical light curve obtained in section \ref{sect:lcana}, which we have also normalized to the X-rays for data after 5 ks, i.e., excluding the first {\it Swift} orbit, which clearly shows that these early data are offset from the fit (see Sect.\ref{sect:SED}). 
}
\label{fig:lcxo}
\end{center}
\end{figure}

\begin{table*}[t]
\centering
\caption{SED Fit Results}
\begin{tabular}{l c c c c c c c}
\toprule
 Epochs & Model$^c$ &$\chi^2/$d.o.f.	&  $\beta_{low}$ 	  & $\beta_{high}$ & $E_{break}$	 &$N_\mathrm{(H,int)}$$^e$  & $E(B-V)$$^e$	\\
(day)&      --           &   --                 &   --   & --        & (keV) & ($10^{21}\,{\mathrm cm}^{-2})$ & (mag) \\
\midrule
0.03$^a$& BPL  &  1.10      & $0.631_{-0.005}^{+0.005}$ &  $\beta_{low}+0.5$  &  $2.09_{-0.05}^{+0.08}$   & 1.14 & 0.012 \\
0.10$^a$& BPL  &  1.10      & $0.592_{-0.006}^{+0.008}$ & $\beta_{low}+0.5$ &  $2.63_{-0.33}^{+0.42}$  &  1.14    & 0.012\\
0.03$^b$& PL  &  2.3      & $0.59_{-0.004}^{+0.006}$  &  -- &  --  & 1.14 & 0.012 \\
0.10$^b$& PL  &  1.4      & $0.62_{-0.006}^{+0.006}$ &  -- &  --  & 1.14 & 0.012 \\
\midrule
0.24--1.50$^d$& PL   & 0.99      & $0.705_{-0.005}^{+0.004}$ &   --       &   -- & $1.14_{-0.28}^{+0.26}$ & $0.012\pm0.009$\\
\bottomrule
\end{tabular}
\begin{tablenotes}\footnotesize
\item $^a$ These two SEDs share extinction and dust absorption from later epochs, but the spectral-model parameters are left free. 
\item $^b$ Like $^a$ but using a single power-law model. 
\item $^c$ We use two models: a broken power law (BPL) and a power law (PL).
\item $^d$ Joint fit: these two SEDs share the same model, and only the normalization is left free.
\item $^e$ The first 4 SEDs have the same $N_\mathrm{(H,int)}$ and $E(B-V)$, fixed to the values found in the late-time fit.
\end{tablenotes}
\label{tab:aftsed}
\end{table*}

\subsection{The early Optical to X-ray spectral energy distribution}
\label{sect:SED}

We used \texttt{Xspec v12.13.0} \citep{Arnaud1996a} to
simultaneously model the optical and X-ray spectral
energy distribution (SED) at the logarithmic mean time of 0.03, 0.1, 0.24  and 1.5 days, corresponding to the first, second, and fourth \swift/XRT observing epochs, and an additional late time epoch before the rise of the SN and corresponding to the third VLT/X-shooter spectra. 
They correspond also to 2 epochs before and 2 epochs after the break in the light curve (at about 0.1-0.2 days, see Sect. \ref{sect:lcana} and Table~\ref{tab:lcfitpara}). 
To build these SEDs, we first created the XRT spectra using the time-slice tool in the XRT repository \citep{Evans2007a,Evans2009a}. We then shifted the optical data closer to these times using the decay indices found above.
For all epochs we applied a Galactic equivalent hydrogen column density of $N_H=3.16\times10^{21}\, \textnormal{cm}^{-2}$ \citep{Willingale2013a}
and a foreground Galactic dust extinction (Sect.~\ref{sec:dataphot}). The redshift was fixed to $z=0.117$.
For the last epoch at 1.5 days, we used the pure AG magnitudes obtained from the procedure described above (Sect. \ref{sect:lcana}).\footnote{The difference is $g=+0.15$, $r=+0.14$ and $z=+0.07$ mag.}

We started analyzing the data from the last 2 epochs, to avoid possible degeneracy between the fireball model breaks in the SED and the absorption. At late times the breaks are commonly far from the optical and XRT bands, and after the jet break they should not shift in time, according to the standard fireball scenario. 
These late-time fits show that a single power-law fits both epochs well.
Therefore, it is reasonable to assume that these two epochs share the same physical synchrotron emission model. 
We jointly fit both epochs with the same shared parameters and we find the best fit ($\chi^2/d.o.f.=853.2/865=0.99$) for a common spectral slope of $\beta_{\rm opt}=\beta_{\rm X}=0.705_{-0.005}^{+0.009}$, with intrinsic $N_H=1.14_{-0.28}^{+0.16}\times10^{21}\,  \textnormal{cm}^{-2}$ and negligible intrinsic dust extinction. This is in particular best modelled with $E(B-V)=0.012\pm0.009$ mag and the LMC extinction-law \citep[][compared to SMC- or MW-laws]{Pei1992a}, which also explains the possible 2175~\AA{} absorption feature visible in the $uvm2$-band in the earliest SEDs 
(see Sect.~\ref{sec:dust} and Fig.~\ref{fig:sedopt}).

Afterwards, we fitted the first epochs at 0.03 and 0.1 days, keeping the intrinsic absorption and dust extinction fixed at the values found above (letting them vary produces consistent results).
We find that the first 2 SEDs (0.03 and 0.1 days) are best modelled by a broken power-law with $\beta_{opt}\sim0.6$, and $\beta_{X}=\beta_{opt}+0.5$, which gets harder with time and with a spectral break at $\sim2$--$2.5$~keV in both epochs ($\chi^2/d.o.f.=1345.8/1220=1.10$). 
The lower-frequency branch of the spectral slope is shallower than the index found above for the 2 later epochs ($\sim0.6$ vs. $\sim0.7$). 
This behaviour is also evident from a simple look at the light curve: at 0.03 days the X-rays are brighter than a simple extrapolation of the optical light curve (see Fig.~\ref{fig:lcxo}; dotted line). Another interesting feature is the spectral break that could shift from $2.09_{-0.05}^{+0.08}$ to $2.63_{-0.33}^{+0.42}$ keV. 
We can interpret this as a synchrotron emission cooling break shifting to higher frequencies \citep[e.g.,][]{Sari1998a}, as expected for a wind environment rather than a constant interstellar medium (ISM) within the fireball model \citep[e.g.,][]{Chevalier1999a}.
However, at 0.1 days, the uncertainty in the break time is large, making it impossible to precisely determine how fast it evolves and confirm this scenario. Moreover, this feature is not commonly seen in afterglows \citep{Schulze2011a},
and it can also arise from a different mechanism that enhances the early X-rays.
In the following, we adopt an ISM scenario but also report wind-model results for completeness. 
We summarize the results of the fits in Table \ref{tab:aftsed}.

These results agree with past findings on GRB afterglows \citep{Ronchini2023a}.
This event is
among the 19 out of 30 GRBs analyzed in that work, for which the shallow-phase has an optical-to-X-ray spectrum fully consistent with synchrotron emission from a single population of shock accelerated electrons \citep[Sample 1 in][]{Ronchini2023a}. The mean X-ray luminosity during this phase is $log(L_X)=45.76^{+0.09}_{-0.11}$ erg s$^{-1}$ 
and its duration is $t_\mathrm{b,X}=21.2$ ks. Therefore, we infer a radiated isotropic energy of $E_\mathrm{X,iso}=1.3\times10^{50}$ erg. 
By assuming a radiative efficiency of $\eta\sim0.1$
in agreement \citep[e.g.,][]{Beniamini2016a}, we conclude that the bolometric isotropic kinetic energy of the jet is at least $E_\mathrm{k,iso} \sim 1\times10^{51}$ erg\footnote{We note that GRB efficiency estimates range between 0.1 \citep[][]{Davanzo2012a} and 0.9 \citep[e.g.,][]{Zhang2007a}. However, a reassessment by \citet{Beniamini2015a, Beniamini2016a} indicates typical efficiencies of 0.1--0.2.}.

\subsection{Collimation-corrected energy \label{sec:collene}}

Under the assumptions of the scenario depicted above, we derive the half-opening angle of the jet.
Assuming that the light curve break at $0.215$ days (Tab. \ref{tab:lcfitpara}) represents the jet break (see Sect.~\ref{sect:lcana}), we use it to measure the collimation of the jet \citep{Sari1999a}.
Following \citep[][]{Frail2001a} we calculate this angle using the following equation for a uniform jet expanding in a constant-density medium:
\begin{equation}\label{angleism}
\begin{aligned}
    \frac{\theta_\mathrm{ISM}}{\mathrm{rad}}
    =0.057~\left(\frac{E_{\gamma, \rm iso}}{10^{53}\,{\rm erg}}\right)^{-1/8}    \left(\frac{n}{\rm 0.1~cm^{-3}}\right)^{1/8} \\
     \times \left(\frac{\eta}{0.2}\right)^{1/8} ~
    \left(\frac{t_{\rm jet}}{\rm 1~day}\right)^{3/8} ~
    \left(\frac{1+z}{2}\right)^{-3/8} \,,
\end{aligned}
\end{equation}
\noindent where $E_\mathrm{tot,iso} =E_\mathrm{\gamma,iso}$ (Sect.~\ref{sec:grbene}, $n=1\,\rm cm^{-3}$ is the number density of the medium, $\eta=E_\mathrm{\gamma, iso}/E_\mathrm{tot,iso}=0.2$\footnote{Thus $E_{\rm kin,iso}=(1/\eta -1)\,E_{\gamma,\rm iso}$} is the typical radiative efficiency assumed for this calculation, $t_{\rm jet}$ is the jet-break time, while $z=0.117$ is the redshift of the event. 
We find $\theta_{ISM}=0.084\pm0.003$ rad ($4.8\pm0.2$ deg), in agreement with other events \citep[e.g.,][]{Laskar2014a,Laskar2018a,Rossi2022b}.\footnote{A similar, just slightly larger value is obtained using \citep{ZhangMacFadyen2009a}.}

Using the above efficiency, we obtain $E_\mathrm{k,iso}=10^{52}$ erg, consistent with the lower limit of $E_\mathrm{k,iso} > 1\times10^{51}$ erg reported at the end of Sect.~\ref{sect:SED}.
If we consider the outflow to be collimated, the
`true' gamma-ray energy of the jet is
$E_{\gamma}=E_\mathrm{\gamma,iso} ~(1-\cos(\theta_{\rm jet}))
\simeq 9\times10^{48}$~erg.
Therefore, 
we estimate the `total collimated energy' of the jet to be $E_\mathrm{tot} \simeq E_\mathrm{\gamma}/\eta \simeq 0.4\times10^{50}$~erg. 
Assuming a wind medium, 
and using the following equation from \citep[][]{Bloom2003a}:
\begin{equation}\label{anglew}
\begin{aligned}
    \frac{\theta_\mathrm{wind}}{\mathrm{rad}}
    =0.169~\left(\frac{E_{\gamma, \rm iso}}{10^{52}\,{\rm erg}}\right)^{-1/4} 
     A_\star^{1/4}~
    \left(\frac{t_{\rm jet}}{\rm 1~day}\right)^{1/4} ~
    \left(\frac{1+z}{2}\right)^{-1/4} \,,
\end{aligned}
\end{equation}
\noindent we obtain 
$\theta_{wind}=0.126\pm0.007$ rad ($7.2\pm0.4$ deg) 
and thus
$E_\mathrm{tot}  \simeq 0.9\times10^{50}$~erg.


\begin{figure}[!tp]
\begin{center}
\includegraphics[width=0.5\textwidth,angle=0]{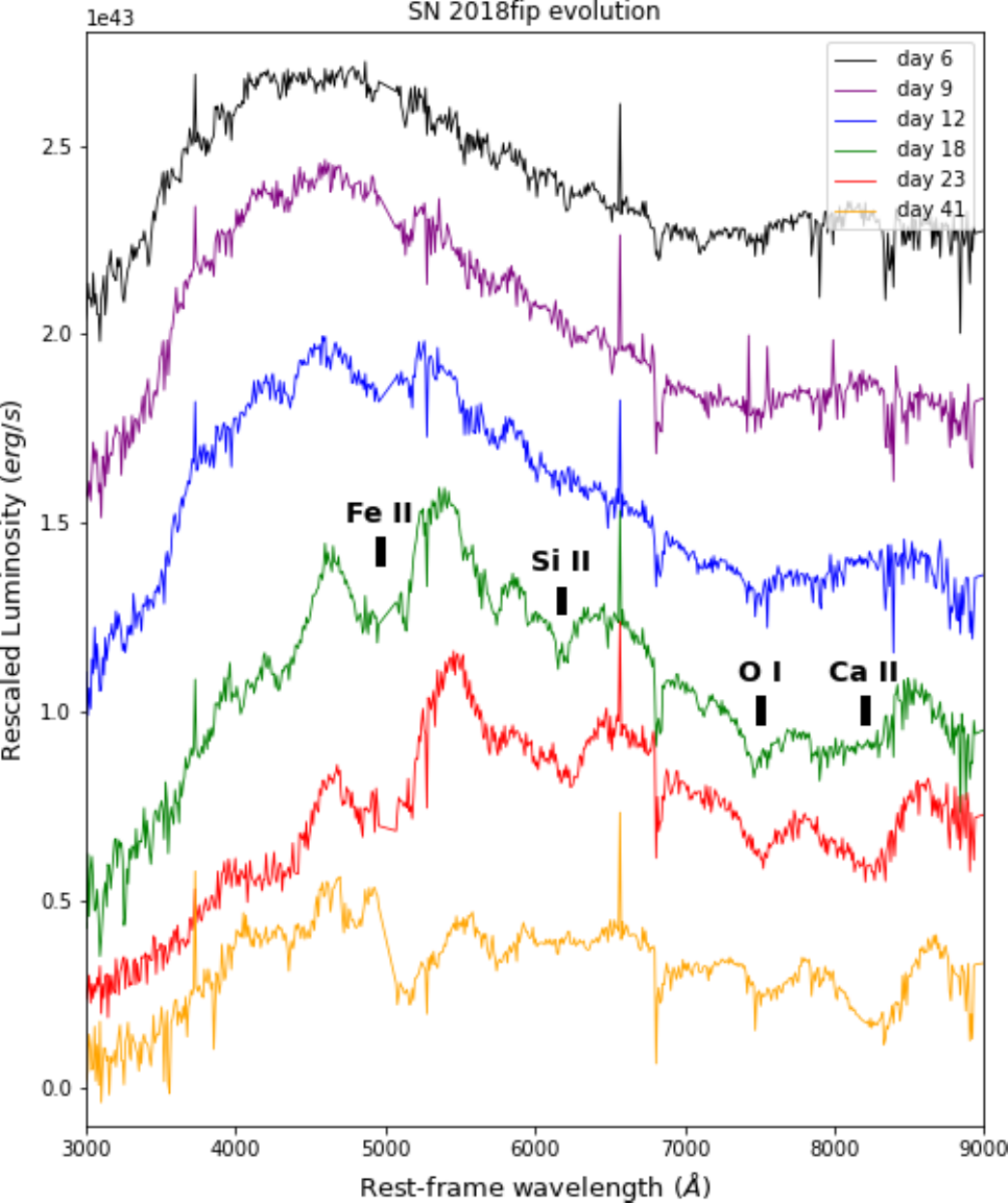}
\caption{Evolution of the spectra of SN 2018fip in rest-frame. 
The absorption features marked in the image refer to the spectrum obtained $\sim$5 days after the peak (day 18). Phases are in the observer frame.}
\label{fig:snseq}
\end{center}
\end{figure}

\subsection{Classification and bolometric light curve of the SN} \label{sec:bolo}


To classify SN 2018fip, we compared our X-shooter spectra of SN 2018fip at 12, 18, and 23 days with other SNe using SNID \citep{Blondin2011a}. We find that the broad-line features most closely match those of the broad-lined Type-Ic SN 2002ap at a compatible phase and the same redshift. For example, at 12 days the best match is obtained for $-2.4$ days relative to maximum light.

After separating the afterglow and SN contributions from the data as explained in section \ref{sect:lcana}, we constructed a bolometric light curve of the supernova emission using our $g'r'i'z'$J photometry. 
We analyzed the light curve using the analytical model developed by \citet{Arnett1982a} for
Type-Ia SNe, which can also be applied to core-collapse Type-Ic SNe with a few careful considerations : i)
the model assumes spherical symmetry for the SN ejecta; ii) it assumes that the total amount of nickel is concentrated at the centre of the ejecta.
These assumptions may not be appropriate in the case of highly rotating progenitor stars and asymmetric explosion \citep{Cano2017a,Izzo2019a}, in which case the derived parameters can change significantly \citep[e.g.,][]{Dessart2017a}.
The model provides an estimate of the total kinetic energy of the SN ejecta, given the expansion velocity measured from P-Cygni absorption of spectral features around the peak brightness of the SN, as well as an estimate of the total amount of nickel synthesized in the explosion and the optical opacity. 
In particular, we derive the expansion velocity from the blueshift of the \ion{Fe}{ii} multiplets, along with the \ion{Si}{ii} $\lambda$6355 doublet, \ion{O}{i} $\lambda$8446, and \ion{Ca}{ii} triplet $\lambda$8492 absorption features (Fig.~\ref{fig:snseq}), fitting each with a Gaussian profile, similarly to \citet{Modjaz2016a}.
From the spectrum of the SN on rest-frame day 14 (12 days in observer frame), we infer an expansion velocity of $v_\mathrm{ej} = 15000 \pm 1000$ km s$^{-1}$. With this assumption, the best fit to the bolometric light curve (Fig.~\ref{fig:bolo}) with the Arnett model gives a total ejected
mass in the SN of $M_\mathrm{ej} = 2.4^{+1.7}_{-0.7}$ M$_{\odot}$, a kinetic energy of $E_\mathrm{k,SN} = (3.2^{+2.3}_{-1.0}) \times
10^{51}$ erg, and a synthesized nickel mass of $M_{Ni} = 0.19^{+0.15}_{-0.09}$ M$_{\odot}$, with a resulting opacity of $k = 0.05 \pm 0.02$ cm$^2$g$^{-1}$.
We find that SN~2018fip reaches a peak luminosity of $(5.70 \pm 0.41) \times 10^{42}$~erg~s$^{-1}$ at $14.9\pm0.17$ days after the GRB trigger (rest-frame).

\subsection{Spectral modelling of the SN} \label{sec:snmod}

\begin{figure}[!tp]
\begin{center}
\includegraphics[width=0.49\textwidth]{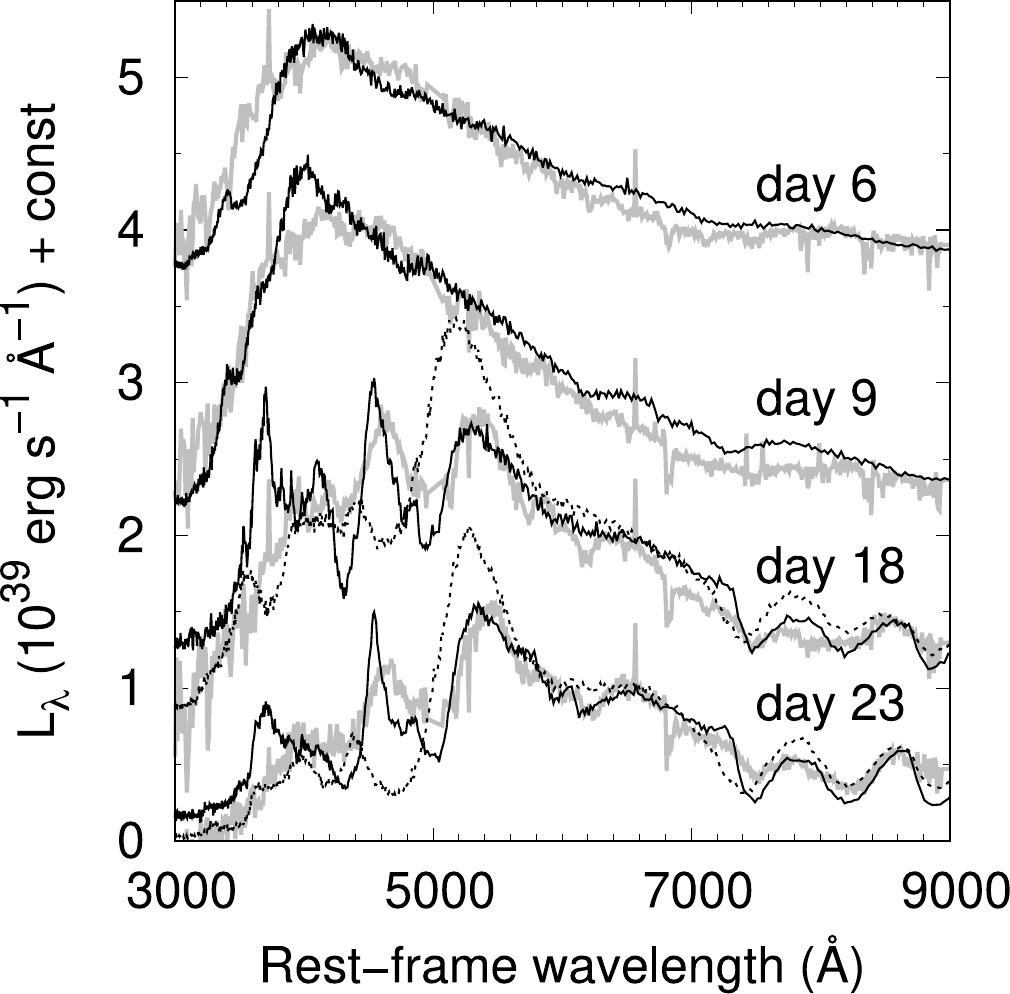}
\caption{Synthesized spectra as compared with those of SN 2018fip. Black-solid lines represent the reference model spectra. The black-dashed lines (on days 18 and 23) show the synthesized spectra with the high-velocity component included in the spectral synthesis calculations, demonstrating that this component is not present at late epochs. The observed spectra are shown with grey lines. The same constant in flux is added for the model and data as an offset for presentation purpose (3.6, 2.0, 0.8 and 0.0 for day 6, 9, 18 and 23, respectively)).}
\label{fig:spec_ref}
\end{center}
\end{figure}

\begin{figure}[!tp]
\begin{center}
\includegraphics[width=0.5\textwidth]{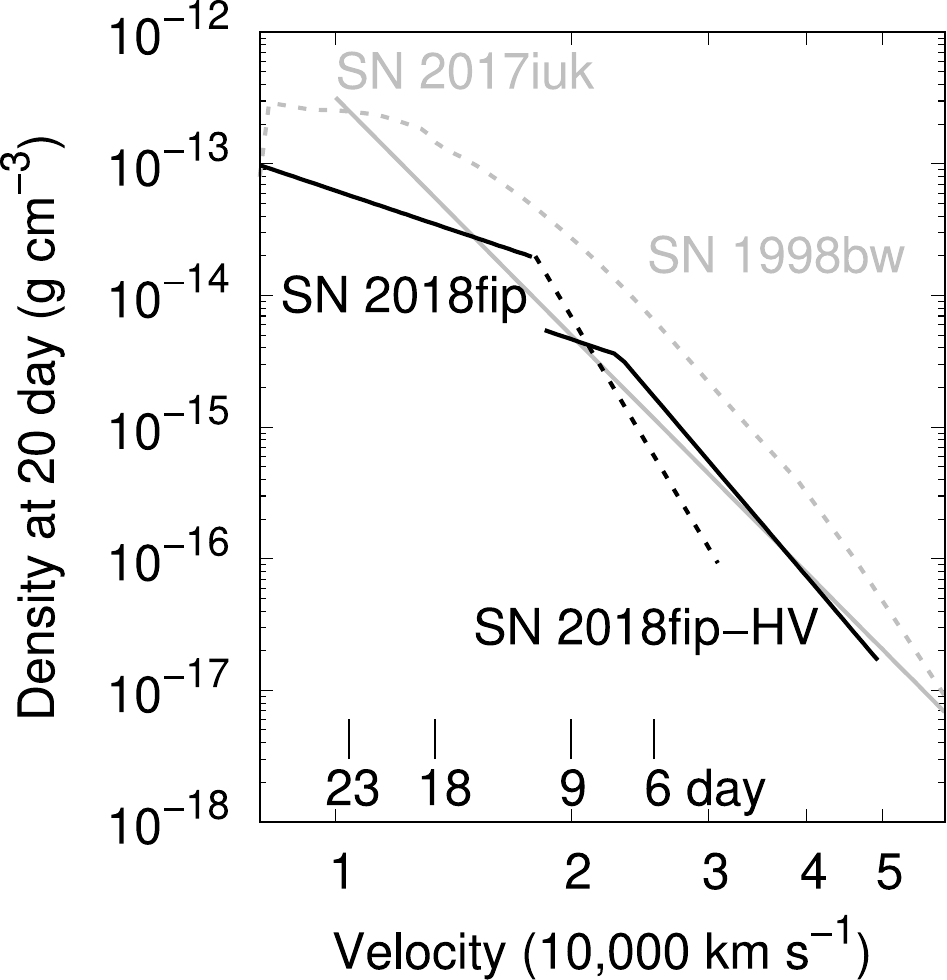}
\caption{The input density structure for the reference spectral model (black-solid lines); the inner lower-velocity component for the spectra on day 18 and 23 (`SN 2018fip') and the outer high-velocity component for day 6 and 9 (`SN 2018fip-HV'). The black-dashed line shows the upper limit set by the spectral synthesis analyses on days 18 and 23 (see the main text). The density structures derived for GRB-SNe 2017iuk \citep{Izzo2019a} and 1998bw \citep{Iwamoto1998} are shown by grey-solid and grey-dashed lines, respectively.}
\label{fig:ejecta}
\end{center}
\end{figure}

To remove the contribution of the afterglow and host from the spectra of SN 2018fip, we used our light-curve modelling to construct $g^\prime r^\prime i^\prime z^\prime J$ SEDs at spectroscopic epochs after day 6 and subtracted them from the dereddened spectra.
Figure \ref{fig:snseq} shows the spectral evolution of SN 2018fip. 
We note that \citep{Buckley2018atel} initially claimed an emerging SN based on spectroscopic data obtained 27.2 h after the burst with the SALT telescope.
However, our second spectroscopic observation with VLT/X-shooter obtained 35.5 h (1.48 days) after the GRB \citep{Heintz2018GCNa} does not show evidence for these broad undulations or deviations from a single power-law with spectral index $\sim0.7$.
The spectra remain featureless until the SN light-curve peak (day 12), when some absorption features start to appear.
We performed spectral synthesis calculations 
using TARDIS \citep{Kerzendorf2014}. Our reference model is shown in Figure \ref{fig:spec_ref}. The input density structure for the reference model is shown in Fig. \ref{fig:ejecta}. 

\begin{figure}[!htp]
\begin{center}
\includegraphics[width=0.5\textwidth]{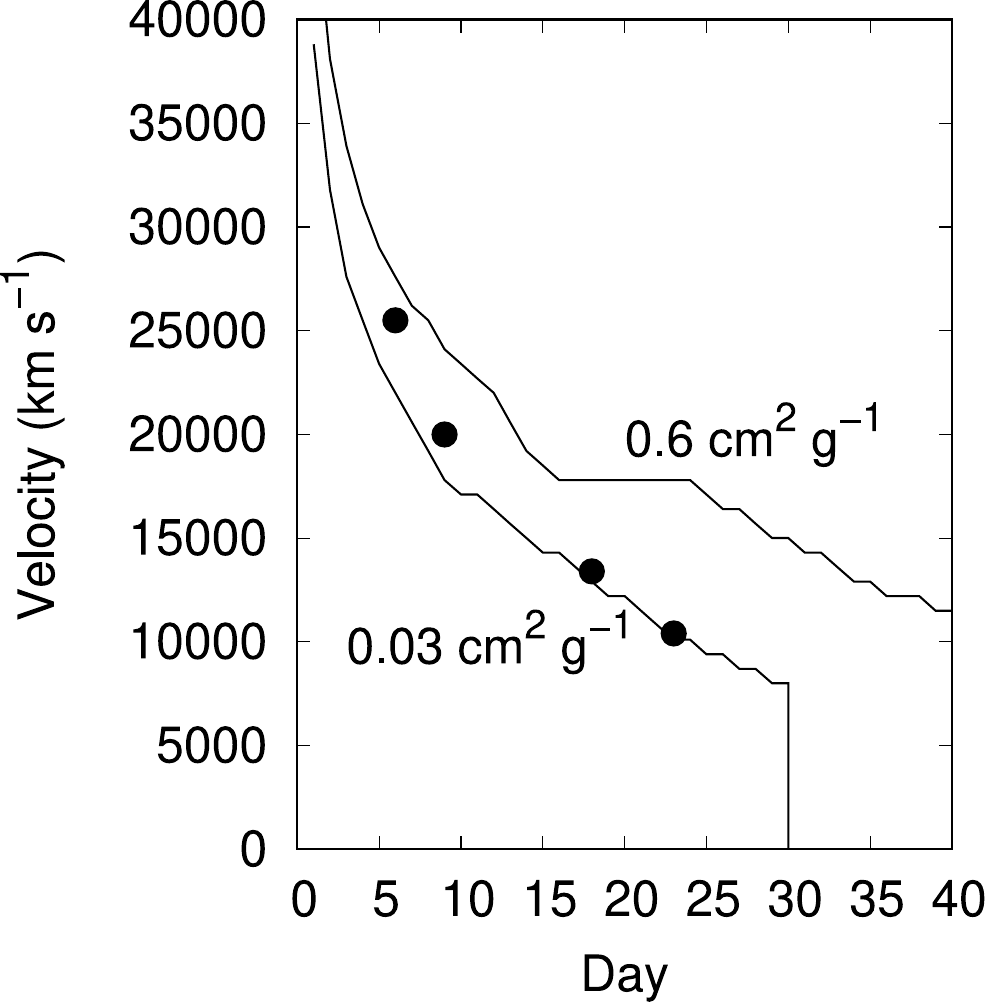}
\caption{The photospheric velocity used in the reference model (points). The two solid curves show the expected evolution of the photospheric velocity calculated for the reference density structure adopting a constant opacity of $0.03$ or $0.6$ cm$^{2}$ g$^{-1}$;
in the TARDIS spectral-synthesis simulations, the input photospheric velocity is varied as a parameter within 
the range between the two lines shown here (see Sect.~\ref{sec:snmod}). 
}
\label{fig:vel}
\end{center}
\end{figure}

To study the temporal evolution, we started with the two spectra after SN peak observed on days 18 and 23. We first assume a density structure described by a single power law. 
We roughly constrain the photosphere position (i.e., set the photospheric velocity as the inner boundary for spectral synthesis) by requiring that the optical depth integrated back from the outer region to the photosphere is about unity for the putative opacity in the range of 0.03 - 0.6 cm$^{2}$ g$^{-1}$ that covers the C+O-rich to Fe-rich composition \citep{Mazzali2001}. Then, the photospheric velocity is varied within this range (Fig.~\ref{fig:vel}), as well as the composition in the spectrum synthesis simulations. 
For the composition, we assume, for simplicity, that the relative mass fractions among Fe-peak elements produced by complete Si burning are universal, guided by typical results from explosive nucleosynthesis calculations \citep[e.g.,][]{Maeda2002}; $X(^{56}{\rm Ni}):X({\rm Ni}):X({\rm Fe}):X({\rm Co})=0.1:0.013:0.01:0.007$, where the normalization here is set by the homogeneous mixing of all $^{56}Ni$ ($\sim 0.4\,M_\odot$) over the entire ejecta ($\sim 4\, M_\odot$). The same is the case for Ti, Cr and Ca created by the incomplete Si burning; $X({\rm Ti}):X({\rm Cr}):X({\rm Ca})=7 \times 10^{-3}:2 \times 10^{-3}:0.01$. We vary the total abundance of these two element groups independently as parameters in the spectral synthesis. The remaining mass fractions are set to be those in a  typical C+O-rich layer; a mixture of C, O, Mg together with a half of the solar abundance for the heavier elements as the progenitor abundance.
We also varied the mass fractions of Si and S, but adding these elements above the progenitor values produced overly strong absorption features. Therefore, we set their additional contribution from explosive burning to zero for these elements; this is not unexpected if the explosive nucleosynthesis takes place following a jet-like explosion \citep[e.g.,][]{Maeda2003}. 
In summary, for a given density structure, we have three main parameters: the photospheric velocity, the mass fraction of $^{56}$Ni (representing complete Si-burning products), and the mass fraction of Ca (representing incomplete Si-burning products).

After tuning these parameters to roughly match the observed spectra at days 18 and 23, we obtain our `final' reference model for the low-velocity part of the ejecta up to $\sim 20,000$ km s$^{-1}$. 
We favour a shallow density distribution below $\sim 20,000$ km s$^{-1}$ ($\rho \propto v^{-2}$ in the reference model) for two reasons. First, reproducing the spectral evolution between days 18 and 23 requires a rapid decrease in the photospheric velocity. Second, the shallow density distribution reproduces line profiles that roughly match the observed spectra. 
As seen in Fig. \ref{fig:vel}, the photospheric velocities used in the model on day 18 and 23 satisfy the constraint from the optical depth. For the composition, our final values are $X(^{56}{\rm Ni}) = 0.15$ and $X({\rm Ca}) = 0.003$ below $15,000$ km s$^{-1}$ while $X(^{56}{\rm Ni}) = 0.03$ and $X({\rm Ca}) = 6 \times 10^{-4}$ above it (up to $\sim 20,000$ km s$^{-1}$). 
Although we do not aim to derive the composition structure accurately, the mass fractions of heavy elements appear to increase toward lower-velocity material, as expected from typical one-dimensional explosion simulations \citep[e.g.,][]{Nomoto2006}. The mass fractions of the Fe-peak elements in the inner region ($ 15,000$ km s$^{-1}$) are roughly the values expected from homogeneous mixing. 

We explored spectral synthesis simulations to identify ejecta structures that  produce synthetic spectra roughly matching the spectra at the earlier epochs (days 6 and 9), independent of the structure applied to the later epochs. These spectra show strong suppression below $\sim 4000$ \AA~ and a largely featureless continuum. 
We explain these two features by a photosphere formed in a high-velocity, Fe-rich outer region ($> 20,000$ km s$^{-1}$); in the reference model (Fig. \ref{fig:spec_ref}), we set the $^{56}$Ni mass fraction to $X(^{56}{\rm Ni}) = 0.24$.
The lower $^{56}$Ni content or lower photosheric velocity results in too blue a continuum without sufficient absorption in the blue portion of the spectra. One could keep the same $^{56}$Ni mass density in order to roughly create a similar amount of blue absorption by simultaneously decreasing X($^{56}$Ni) and increasing the density. However, this conflicts with the constraint on the photospheric velocity we set in our model construction (i.e., Fig. \ref{fig:vel}). 

In the high-velocity region, we include stable Ni (i.e., $^{58}$Ni) with the relative mass fraction to $^{56}$Ni set the same as applied to the inner region. 
We do not include other Fe-peak elements or Ca in the high-velocity region, because they  characteristic absorption features at different wavelength regions that clash with the observed featureless spectra. This indicates that this high-velocity component may originate in the highest-entropy region in the explosion \citep[i.e., the one created in the deepest ejecta in a spherically symmetric configuration;][]{sato2021}.
The evolution of the photosphere and its velocity can be seen in figures \ref{fig:ejecta} and \ref{fig:vel}, which shows that the photosphere is in the high-velocity region early on, but is below the high-velocity region on day 18. We further discuss the implications in section \ref{sec:asym}.

We note that the inner distribution adopted to model the spectra on day 18 and 23 does not include the high-velocity component used for the spectra on day 6 and 9.
In `spectral tomography' \citep[e.g.,][]{Mazzali2015}, the ejecta structure is reconstructed from the outermost to the innermost regions by modelling the early spectra first, since material above the photosphere can still significantly affect the observations.
Figure \ref{fig:spec_ref} shows that this procedure does not work for SN 2018fip. 
The dashed curves are the synthetic spectra on day 18 and 23 created with the high-velocity outer component, showing several critical problems; the features dominated by Fe II at $\sim 5,000$ \AA\ never match the data, with the pseudo emission peak shifted to the blue as compared to the observed wavelength, and the associated absorption becoming too strong. This indicates that there is too much (relatively low-temperature) Fe-rich material either along/out of the line of sight (for the absorption/emission problem). 
Also, the high-velocity component produces excessive absorption and heating in the outer layer, shifting the O I absorption toward higher velocities.
Further details can be found in section \ref{sec:addsn}.


\begin{figure}[!tp]
\begin{center}
\includegraphics[width=0.5\textwidth,angle=0]{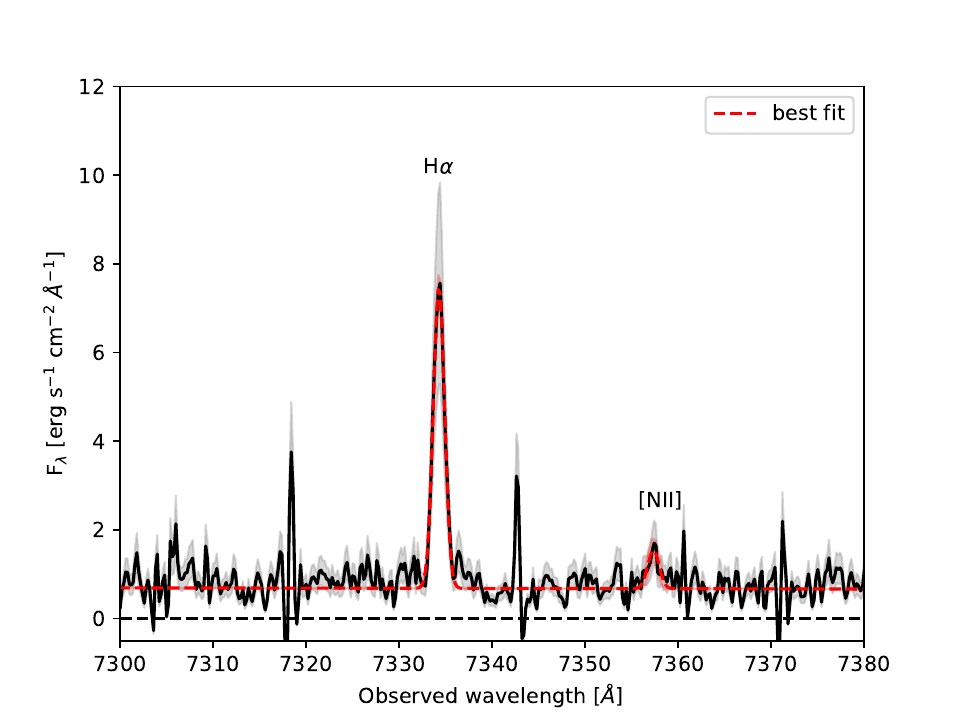}
\caption{
Zoom-in on the H$\alpha$ line and [\ion{N}{ii}] region of the VLT/X-shooter spectrum of the host galaxy 
in units of 10$^{-17}\,$erg$\,$s$^{-1}\,$cm$^{-2}$\AA$^{-1}$.
The detected emission lines are marked.
A fit of the lines and of the continuum is shown in red. 
In grey the error on the spectrum.
}
\label{fig:spectrum} 
\end{center}
\end{figure}

\begin{table}
\centering
\caption{Emission lines and their measured fluxes corrected for Galactic extinction.}
\begin{tabular}{lc} 
\toprule  
Line         &   Flux       \\
              &[10$^{-17}$ erg cm$^{-2}$ s$^{-1}$] \\
\midrule
H$\beta $    $\lambda$4861 & $3.65\pm0.39$  \\
$[$O \textsc{iii}] $\lambda$4959+5007& $5.91\pm0.40$  \\
H$\alpha$    $\lambda$6563 & $9.52\pm0.30$\\
$[$N \textsc{ii}]        $\lambda$6583 & $1.20\pm0.38$  \\
\bottomrule
\end{tabular}
\label{tab:lines}
\end{table}

\begin{figure}[!t]
\begin{center}
\includegraphics[width=0.49\textwidth,angle=0]{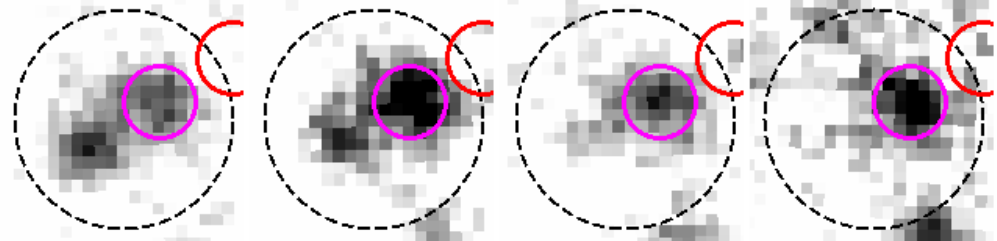}
\includegraphics[width=0.48\textwidth,angle=0]{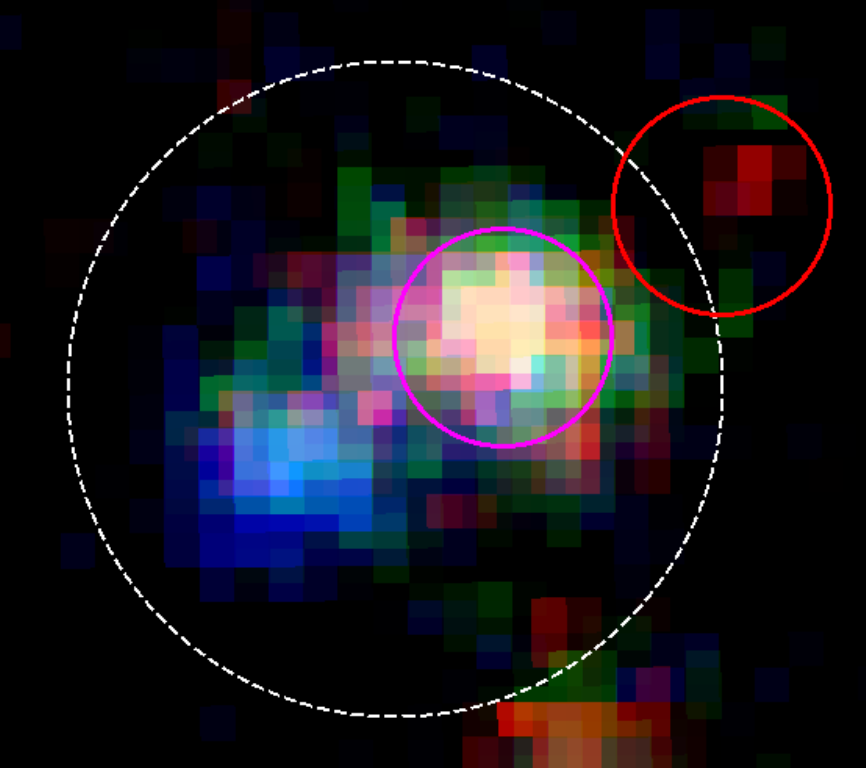}
\caption{$i^\prime,r^\prime\,g^\prime$/RGB-color image of the host galaxy of GRB 180728A. Bluer regions correspond primarily to $g'$-band emission.
The top panels show the individual  
$g^\prime\,r^\prime\,i^\prime\,z^\prime$ GROND imaging. The nearby star (red circle) was  removed using a PSF model, although some residual is left. 
The magenta circle shows the position of the afterglow/SN. The dashed circle has a radius of 1\farcs5 and shows the region used for aperture photometry. North is up, East is left. 
}
\label{fig:host}
\end{center}
\end{figure}

\begin{figure}[thp]
\begin{center}
\includegraphics[width=0.5\textwidth,angle=0]{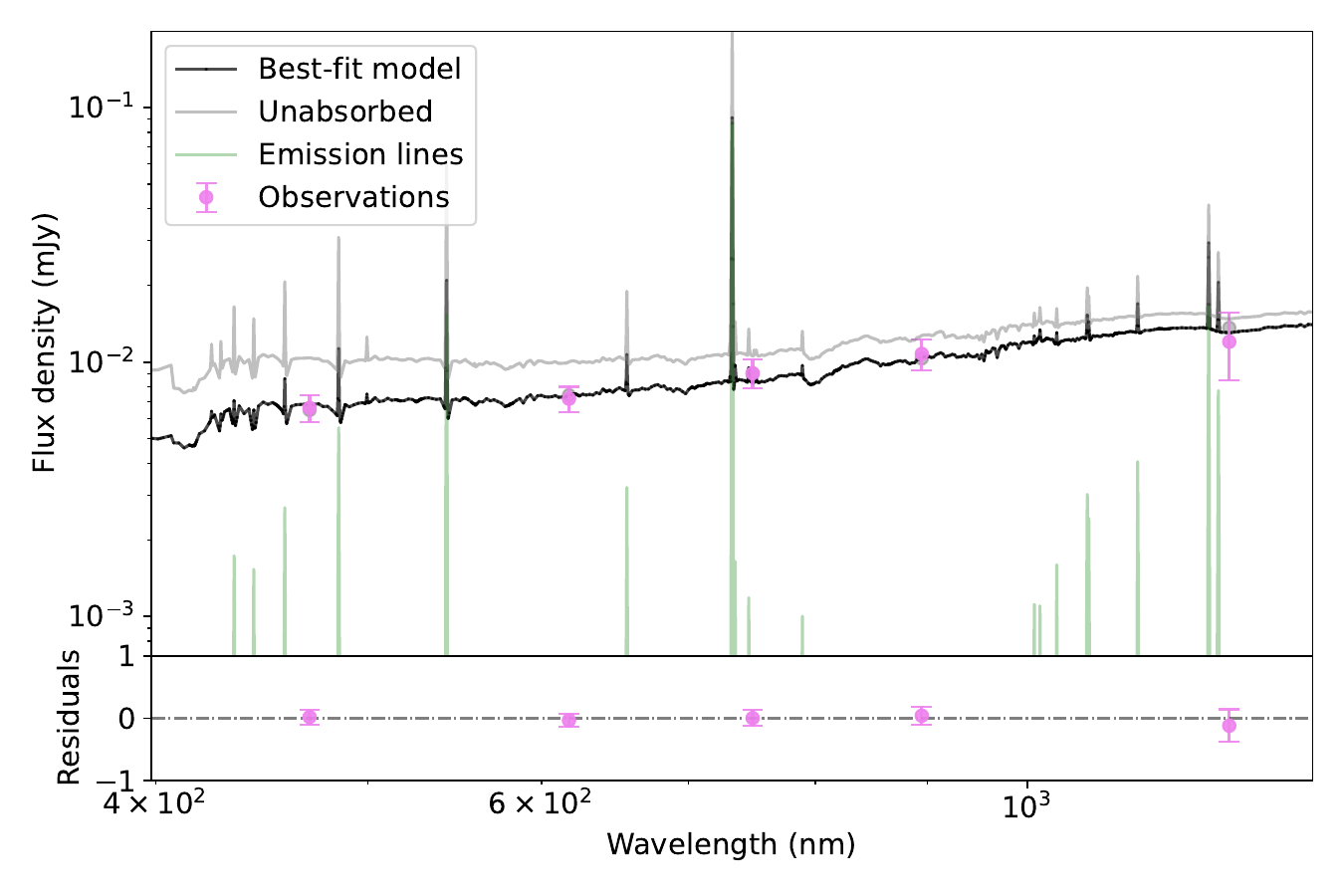}
\caption{The GROND $g^\prime\,r^\prime\,i^\prime\,z^\prime$ and VISTA $J$-band photometry of the host galaxy of GRB 180728A with
{\tt Cigale} (section \ref{sec:host}). The best model (solid line) is shown together with the relative residuals. Photometric measurements are marked by violet empty circles.}
\label{fig:sedhost}
\end{center}
\end{figure}

\subsection{The host galaxy}
\label{sec:host}

Thanks to our late imaging, we can investigate the morphology of the host galaxy. 
We used the last GROND epoch, obtained 254 days after the GRB, which is deeper than the X-shooter images.
Unfortunately, also in this case the presence of the bright star disturbs the analysis.
Therefore, we used the IRAF \texttt{daophot} package \citep{Stetson1987a} to build a PSF model and subtract the nearby star.
After removing the emission from the star, an extended object is clearly visible at the afterglow/SN position in the optical bands (Fig.\ref{fig:host}), which we identify as the host galaxy of GRB 180728A.
The host appears to be made up of two bright regions visible in the $g^\prime$ and $r^\prime$ filters. It is extended in the SE-NW direction with 
major and minor axes of $1\farcs4$ and $0\farcs7$, measured in the $g^\prime$-band.
These correspond to $\sim$3 and $\sim$1.4 kpc at the redshift of the GRB.
The SE region is bluer and contributes approximately half of the host flux in the $g^\prime$ band, indicating a younger stellar population than in the NW region.
Interestingly, the bluer and younger SE-region 
does not coincide with the location of the GRB-SN (color panel in Fig. \ref{fig:host}), although the GRB exploded with a negligible offset from the brightest region, in agreement with the LGRB population
\citep[e.g.,][]{Bloom2002a,Blanchard2016a}.

We obtained aperture photometry within a circular radius of $1\farcs5$  corresponding to 3.3 kpc at the redshift of the burst, large enough to include the entire host complex while avoiding nearby sources.
The galaxy is not detected in the NIR bands with GROND down to the following AB upper limits $J>20.8$, $ H>19.7$, $K_s>17.8$, but it is detected in VISTA $J$ archival images (though not in $K$). We measured the following AB aperture magnitudes, not corrected for the foreground Galactic extinction:
$g^\prime=22.77\pm0.06\,$, $r^\prime=22.40\pm0.06\,$, $i^\prime=21.98\pm0.09\,$, $ z^\prime=21.67\pm0.10$, $J=21.4\pm0.3$. 
To assess whether the late-time data are still affected by SN light, we considered the SN~1998bw light curve, which extends beyond 250 days, together with the best-fit scaling and stretching parameters derived in Section~\ref{sect:lcana}.
We find that the SN should be $g=25.5$, $r=25$, $i=24$, $z\sim24$ which is 2--3 mag fainter than the host. 
 This contribution is included in the photometric error of the host magnitude measurements. 

To study the spectral properties of the host, we used the final spectrum obtained at 71 days, with the slit oriented along the host major axis.
Since there was no detectable continuum in the NIR arm, we limited the analysis to the UVB and VIS arms.
We accounted for the small SN contribution at this late epoch and fine-tuned the absolute flux calibration using late-time $g^\prime$, $r^\prime$, $\it i^\prime$, and $\it z^\prime$ GROND photometry at 254 days (see below), which is free from SN and afterglow emission.
The continuum of the host galaxy is clearly detected between
$0.4\,\mu\textnormal{m}-1\,\mu$m. 
We detect the emission lines [\ion{O}{iii}] doublet and [\ion{N}{ii}], Balmer H$\beta$ $\lambda$4861 and H$\alpha$ $\lambda$6563 (Fig.~\ref{fig:spectrum})
at a redshift of $0.1172\pm0.0001$, consistent with that of the afterglow (Sect.~\ref{subsec:spectra}).
All measured fluxes are reported in Table \ref{tab:lines}.
Since the slit follows the host major axis, it captures emission from both star-forming regions (Fig. \ref{fig:host}).
Following the O3N2 metallicity calibration from \citet{Hirschauer2018a}, we find $12+\log(\textnormal{O/H})
 = 8.57$. The Balmer decrement, H$\alpha$/H$\beta = 2.61\pm0.36$, corresponds to $A_V<0.1$ and therefore to a negligible dust content \citep{Osterbrock1989}.
We measure a low star formation rate (SFR) of $0.02\, M_\odot/\mathrm{yr}$ based on the H$\alpha$ line flux, assuming a \citep{Chabrier2003a} initial mass function (IMF) and following \citet[][]{Treyer2007a}. The low SFR and metallicity are consistent with those observed for LGRB hosts at similar redshifts \citep[e.g.,][]{Kruhler2015a,Vergani2015a,Japelj2016a}.

After correcting for foreground extinction, we modelled the host-galaxy SED with the  \texttt{CIGALE} code \citep{Boquien2019a}, fixing the redshift to $z=0.117$. 
Unfortunately, the photometric uncertainties do not allow us to model the two regions separately.
Given the lack of rest-frame UV photometry, it is also difficult to constrain the SFR. 
We can still estimate the SFR using an alternative approach: in $i^\prime$ the host is brighter than in $r^\prime$ and $z^\prime$-bands, which we interpret as H$\alpha$ emission falling within this band at the redshift of the GRB (Fig. \ref{fig:sedhost}). 
We include this effect in the SED fitting because \texttt{CIGALE} accounts for the contribution of emission lines to the photometry via Kennicutt relations \citep{Kennicutt1998a}, allowing an SFR estimate, although the large photometric uncertainties make this constraint very uncertain.
Assuming a delayed star formation history and a \citet{BruzualCharlot2003a} stellar population with a Chabrier IMF, we obtain a best fit to the photometric data ($\chi^2/d.o.f.=0.1$) for a galaxy with a relatively young (0.3–0.9 Gyr) stellar population, dust attenuation $A_V=0.4$ mag (using the \citealt{Calzetti2000} law), a stellar mass of $10^{7.84\pm0.23}\,{\it M}_\odot$, SFR of $0.28_{-0.27}^{+0.31} \, {\it M}_{\odot}\,{\rm yr}^{-1}$,
i.e., $\lesssim0.6\,{\it M}_{\odot}\,{\rm yr}^{-1}$. With these values we derive a specific star formation rate (SSFR) of $0.1$--$9 \, {\rm Gyr}^{-1}$, well within those measured so far in the LGRB host population \citep[e.g.,][]{Savaglio2009a,Hunt2014a,Perley2013a,Vergani2015a,Japelj2016a,Schulze2018a}. The absolute magnitude is $M_B=-16$. 
We conclude that this is a young star-forming dwarf galaxy similar to other LGRB host galaxies found at low redshift \citep[e.g.,][]{Savaglio2009a,Taggart2021a}.
In particular, its properties are similar to the host of GRB 030329, which is just slightly more luminous ($M_{\rm B}\sim-17$; \citealt{Gorosabel2005a,Sollerman2005a,Savaglio2009a}).


\begin{figure}[th]
\begin{center}
\includegraphics[width=0.49\textwidth,angle=0]{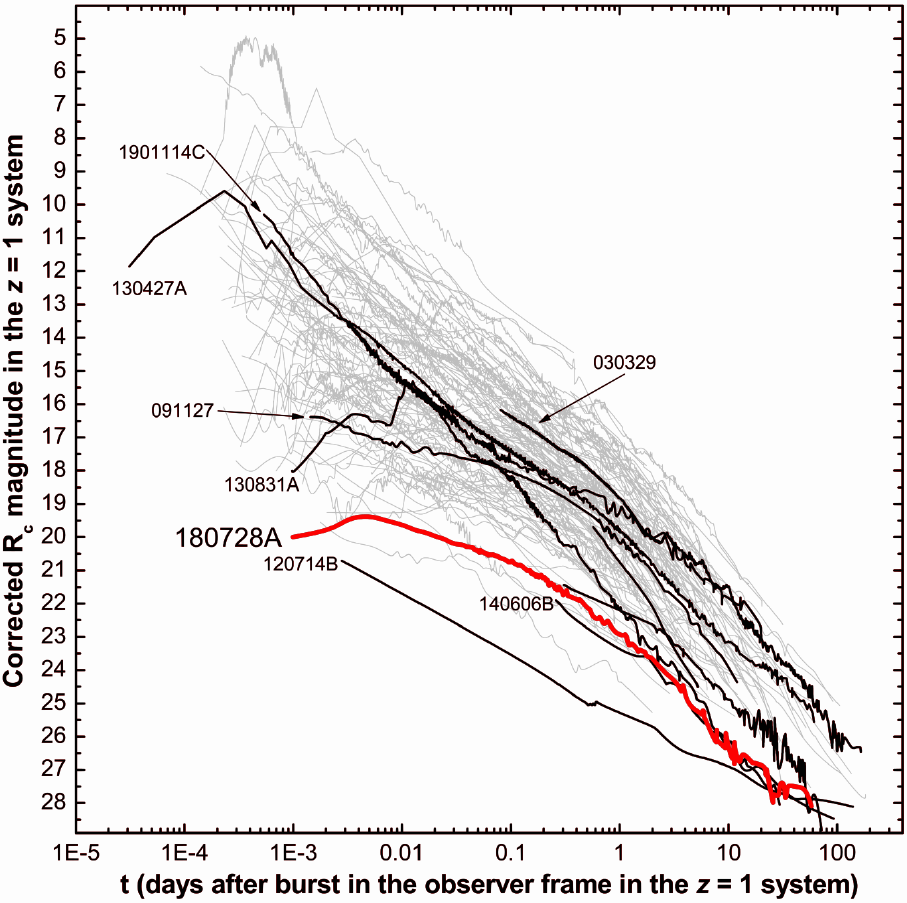}
\caption{Afterglows of GRBs, all transformed to the $z=1$ system following the method of \citep{Kann2006ApJ}. We highlight the afterglow of GRB 180728A, which is seen to be among the faintest known in the context of luminosity, in red. We also mark, with thick dark grey lines, other GRBs at $z<0.5$ that are associated with SNe (the SN emission is subtracted in this plot).}
\label{fig:KannPlot}
\end{center}
\end{figure}

\begin{figure}[th]
\begin{center}
\includegraphics[width=0.5\textwidth,angle=0]{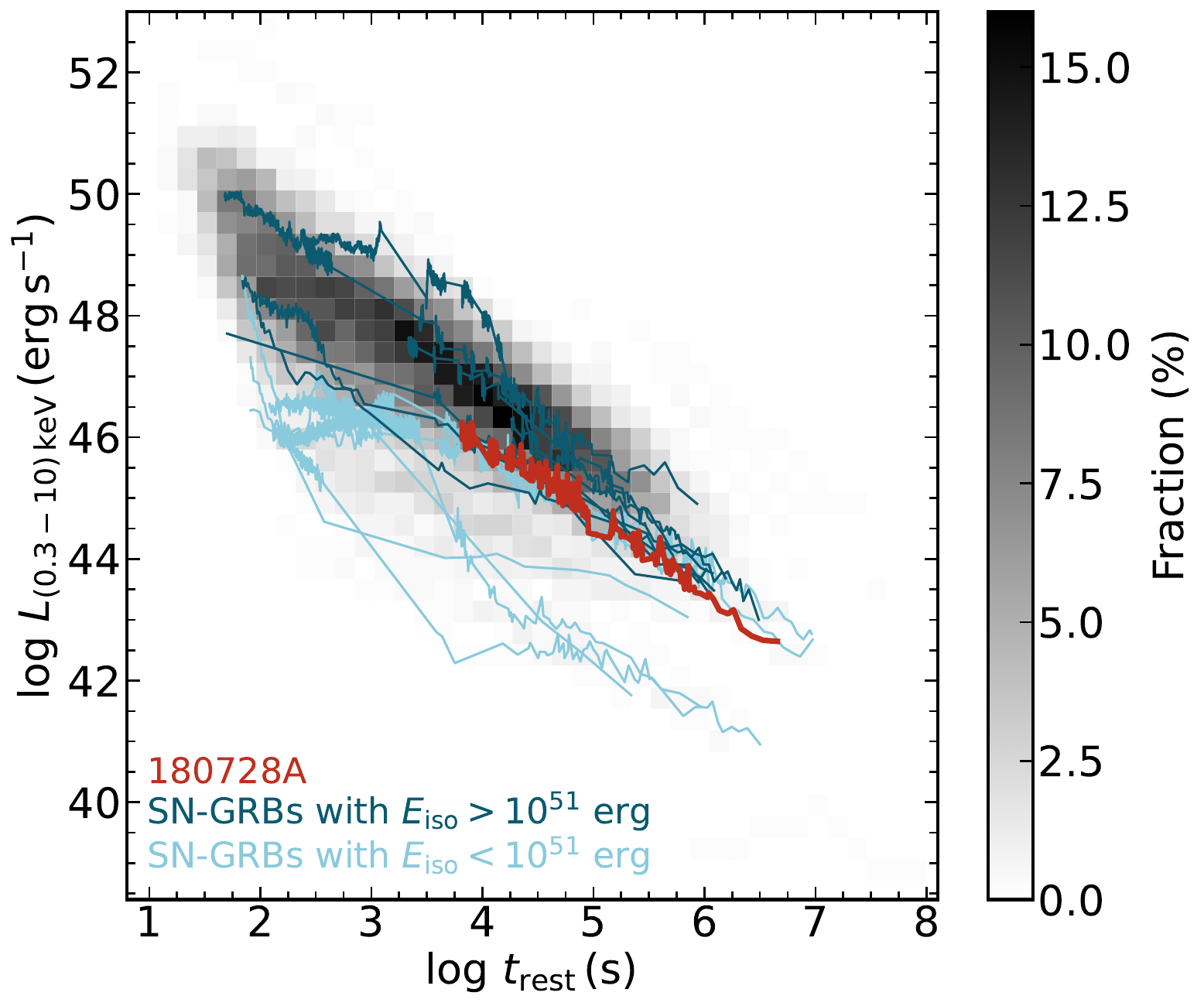}
\caption{
The X-ray light curve of GRB 180728A in the context of the X-ray afterglows of 473 Swift GRBs with known redshifts (until February 2024). For comparison, we overlay the light curves of Swift GRBs with spectroscopically confirmed SNe and divide the sample in 6 SN-GRBs with $E_\mathrm{\gamma,iso}<10^{51}$, and 9 SN-GRBs with $E_\mathrm{\gamma,iso}>10^{51}$ erg. The colour table on the right side translates a grey shade at a given luminosity and time into a fraction of bursts.}
\label{fig:SchulzePlot}
\end{center}
\end{figure}

\section{Discussion}\label{sec:Discussion}

\subsection{The afterglow of GRB 180728A in context} \label{sec:agcon}

Using the redshift and the SED, we transform the pure afterglow light curve to $z=1$ following the method described in \citep{Kann2006ApJ}. Here, 
we correct the afterglow for line-of-sight extinction (negligible in this case) and shift it in time and luminosity to $z=1$, including the $k$-correction. The magnitude offset we find is $dRc=+5.23\pm0.03$. Thereby, it can be directly compared to the large afterglow collection presented by \citep{Kann2006ApJ,Kann2010a,Kann2011a, Kann2024b}. 
We show the comparison in Fig. \ref{fig:KannPlot}, 
with the afterglow of GRB~180728A highlighted in red.
We also highlight with thicker dark gray lines the afterglows of some other GRBs at $z<0.5$ that are associated with GRB-SNe (spectroscopically secure in most cases), where some are labeled. 

Similarly, to place the X-ray emission of GRB~180728A in context, we retrieved X-ray light curves of all \swift\ GRBs up to February 2024 with detected afterglows (at least two epochs) and known redshifts from the Swift Burst Analyser\footnote{\href{https://www.swift.ac.uk/burst_analyser/}{https://www.swift.ac.uk/burst\_analyser/}} \citep{Evans2010a}. 
The density plot in Fig.~\ref{fig:SchulzePlot} displays the parameter space occupied by these 473 bursts (using the method described in \citealt{Schulze2014a}). 
In this case, for comparison, we have also highlighted other GRBs with spectroscopically-confirmed SNe and separated them in low- and high-$E_\mathrm{\gamma,iso}$.\footnote{$E_\mathrm{\gamma,iso}<10^{51}$ erg GRBs: 
060218, 100316D, 120422A, 161219B, 171205A, 190829A; $E_\mathrm{\gamma,iso}>10^{51}$ erg GRBs: 050525A, 081007, 091127, 101219B, 111209A, 140606B, 171010A, 190114C, 230812B} 
From these figures, it can be seen that the optical afterglow of GRB 180728A, while bright observationally, is among the least luminous known. Definite SN-associated GRBs have early afterglows
that range from typical to faint. 
Indeed, several of these are among the faintest afterglows ever followed up to late times; this is a selection bias stemming from the interest in SN follow-up combined with {\it Swift's} capability to probe the UV regime where GRB-SNe are usually strongly suppressed (but see \citealt{Kann2019AA}), revealing the true afterglow evolution.

As in the optical, the X-ray afterglow of GRB~180728A is slightly under-luminous compared to other \swift\ GRBs, although it is not among the faintest, as in three other events the X-ray 
afterglow declines rapidly without the early shallow phase observed here.
However, we see that GRB 180728A lies in between the two samples of high-energy cosmological GRBs and low-energy and low-redshift GRBs, especially within one day. Therefore, in this case, the X-ray afterglow appears to scale with $E_\mathrm{\gamma,iso}$. \citet{Hu2021a} noted that its \swift-XRT light curve shows late-time bumps at 10 keV that are also observed in the cases of GRBs 190829A and 171205A \citep{Izzo2019a}, although bumps are known since at least GRB 030329 \citep[e.g.,][]{Zeh2004ApJ} and are a possible signature of refreshed shocks \citep[e.g.,][]{Granot2003a,Moss2023a}.

In fact, a striking difference from other GRB-SNe at $z<0.5$ is in the slower fading of the afterglow (see Fig. \ref{fig:KannPlot} and \ref{fig:SchulzePlot}) within the first 0.1 days. None of the other low-redshift GRB-SNe show such early shallow fading except for GRB 091127-SN2009nz
 \citep{Filgas2011a}, which has a light curve that is strikingly similar though $\approx10$ times more luminous.
Indeed, both the prompt emission and the SN are more luminous than those of GRB~180728A/SN~2018fip and are more similar to those of GRB 030329/SN~2003dh.
Not surprisingly, at late times ($>1$ day in rest frame) almost all afterglows show a similar decay after the jet break (an exception being GRB 1304027A, see e.g., \citealt{DePasquale2016a}).


\begin{figure}[tp]
\begin{center}
\includegraphics[width=0.45\textwidth,angle=0]{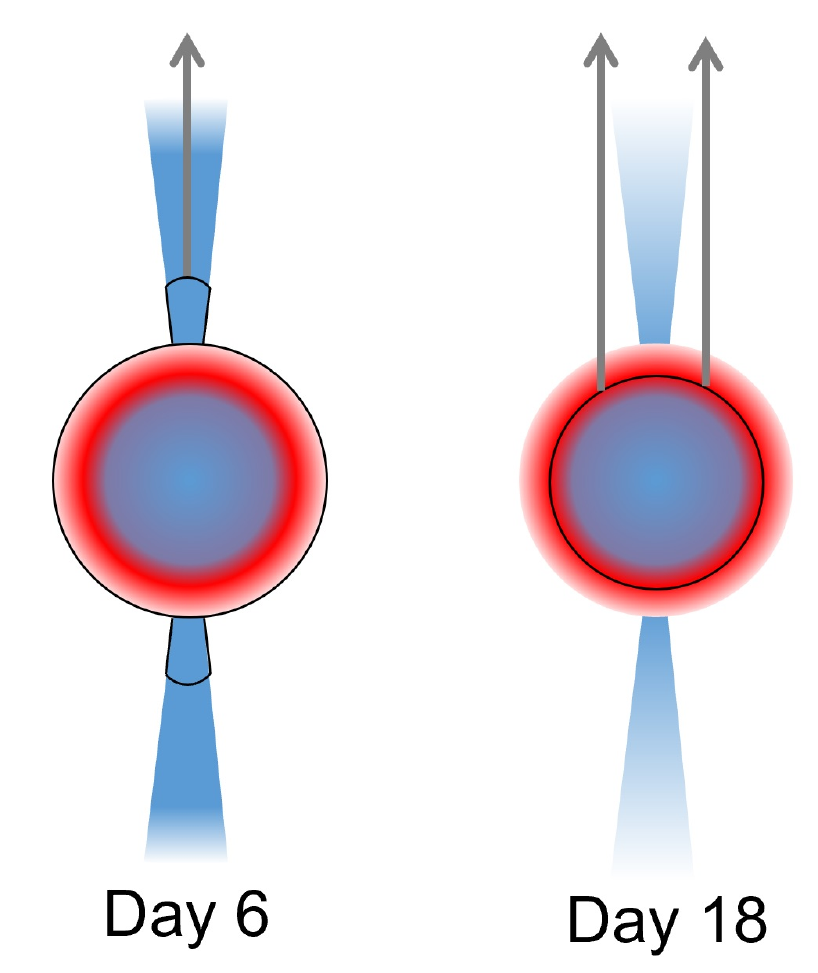}
\caption{A schematic picture for the ejecta structure, showing
a combination of the $^{56}$Ni-rich high-velocity component
confined in a narrow solid angle and the slower quasi-spherical
ejecta. The black line shows the 
photosphere at: i) 6 days,  when it includes the high-velocity high-velocity component;
ii) at 18 days, when the photosphere recedes in the inner ejecta.
}
\label{fig:toy}
\end{center}
\end{figure}

\subsection{On the asphericity and energy of the SN ejecta} \label{sec:asym}

In section~\ref{sec:snmod}, we modelled the spectroscopic data from 6 to 23 days using a high-velocity component for the early phase and a low-velocity component for the late phase. A problem we found is that the high-velocity component must produce detectable signatures at later epochs, but no such traces are seen on day 18 and thereafter.
This discrepancy may have important implications for the ejecta structure, the jet formation, and the explosion mechanism. The most likely cause of the problem in explaining the spectral evolution is the strong assumption of spherical symmetry adopted in TARDIS. In fact, asymmetry/asphericity in the ejecta structure has been extensively discussed for GRB-SNe \citep[e.g.,][]{Maeda2003,Suzuki2022,Maeda2023}. 
One possible configuration that could overcome this problem/difficulty is the one shown in Figure \ref{fig:toy}: a combination of a $^{56}$Ni-rich high-velocity component confined in a narrow solid angle and a slower quasi-spherical ejecta \citep[see also][]{Ashall2019a}. In this configuration, the photosphere is within the high-velocity $^{56}$Ni-rich component in the early phase \citep[when the radiation output can indeed be dominated by this component;][]{Maeda2006}, while it recedes to the inner ejecta component in the late phase. If the solid angle of the high-velocity component is significantly smaller than the size of the photosphere in the late phase, the absorption created in the high-velocity component could be minimized. If we take the characteristic velocity of the high-velocity component as $\sim 20,000$ km s$^{-1}$ and the photospheric velocity on day 23 as $\sim 10,000$ km s$^{-1}$ (Fig. \ref{fig:vel}), this requires that the half-opening angle of the high-velocity component is $<30^{\circ}$, or the fraction of the solid angle ($\Omega/4 \pi$) $<13\%$ assuming the bipolar structure. The high-velocity Fe-rich component may represent a sub-relativistic cocoon associated with a GRB jet \citep{Nakar2015a,NakarPiran2017a,Gottlieb2018a,PiroKollmeier2018a,Izzo2019a}. 

The low-velocity (perhaps quasi-spherical) component has the ejecta mass $M_\mathrm{ej}$, kinetic energy $E_\mathrm{k,SN}$, and Nickel mass $M$($^{56}$Ni) within $\sim 10,000 - 20,000$ km s$^{-1}$ as follows: $\sim 1.5\, M_\odot$, $\sim 3 \times 10^{51}$ erg, and $\sim 0.17\, M_\odot$, respectively. 
Because it does not include material below $\sim 10,000$ km s$^{-1}$ and above $20,000$ km s$^{-1}$, these values provide lower limits. For example, incorporating the density structure above $20,000$ km s$^{-1}$ (dashed line in Fig.~\ref{fig:ejecta}) increases them to $M_{\rm ej} \sim 2, M_\odot$ and $\sim 5 \times 10^{51}$ erg.
On the other hand, the high-velocity component ($> 20,000$ km s$^{-1}$) has {\it isotropic} values of $M_{\rm ej} \sim 0.5 \, M_\odot$, $\sim 3.5 \times 10^{51}$ erg, and $M$($^{56}$Ni) $\sim 0.1\, M_\odot$. Because the high-velocity component occupies a narrow solid angle, the low-velocity component likely dominates the total mass and energy.
A fair comparison to other GRB-SNe (whose values are usually derived  through the Arnett relation) requires care, because asphericity strongly affects the spectral properties of SN 2018fip, particularly its energy.
If we assume a simple spherical scenario, we can sum the inner and outer components obtained above, with an additional hidden internal region of 1 $M_\odot$, and therefore obtain a total of $M_\mathrm{ej}=3\,M_\odot$ and $E_\mathrm{k,SN}\sim6.5\times10^{51}$ erg. Note that the low-velocity ejecta agree well with the values obtained via the bolometric light curve and the Arnett model, probably because both methods assume a homologous expansion with nickel concentrated in the centre . The high-velocity ejecta adds additional energy, and thus the total values are larger than those obtained from the bolometric light curve. 
However, the kinetic energy estimated for the high-velocity component could be overestimated and should be reduced by a factor $<$0.13, considering the half-opening angle of $<30^\circ$ estimated above.


\begin{figure}[!tp]
\begin{center}
\includegraphics[width=0.5\textwidth,angle=0]{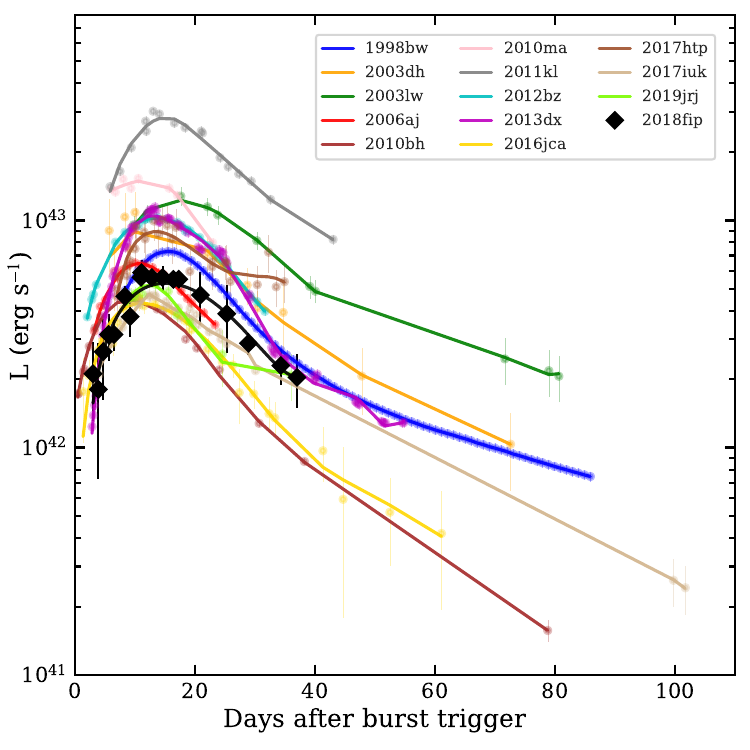}
\caption{The quasi-bolometric light curve of SN 2018fip is compared to those of 13 well-observed GRB-SNe, with low-order spline fits shown for each. Most of the comparison light curves, especially for GRB-SNe observed before 2015, are adopted from the compilation by \citet{Cano2017a}, which incorporates data from multiple earlier studies cited therein. Additional events were included from more recent works by \citet{Cano2017b, Izzo2019a, Melandri2019a, Melandri2022a, Kumar2022a,Kumar2024a}.}
\label{fig:bolocomp}
\end{center}
\end{figure}

\begin{figure}[!tp]
\begin{center}
\includegraphics[width=0.5\textwidth,angle=0]{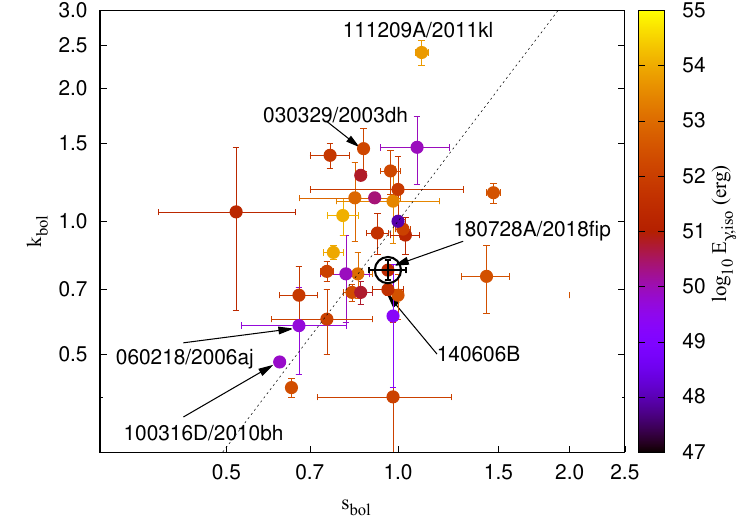}
\includegraphics[width=0.5\textwidth,angle=0]{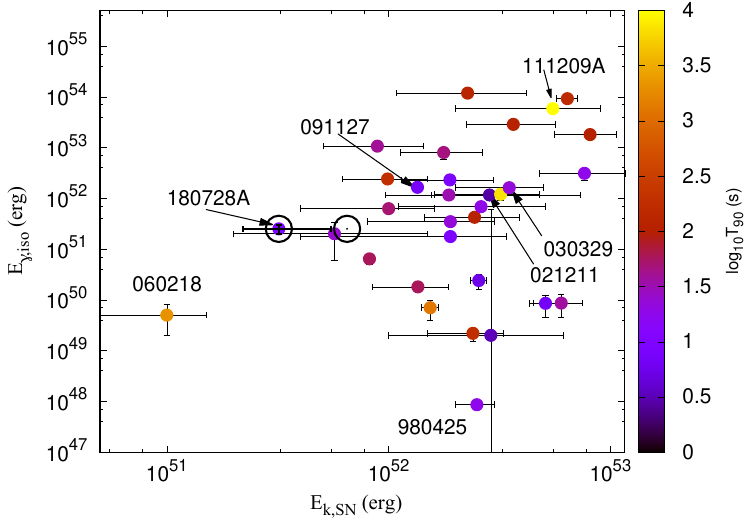}
\caption{ \textit{Top:} Luminosity--stretch diagram color coded by $E_\mathrm{\gamma,iso}$. The dashed line is the best fit from \citep{Klose2019a}.
\textit{Bottom:}
$E_\mathrm{k,SN}$ -- $E_\mathrm{\gamma,iso}$ plane, color coded by $T_\mathrm{90}$. Data is taken from \citep{Cano2017a}, and updated with recent results \citep{Klose2019a,Dainotti2022a,Kann2024a,Dong2025a}. For SN 2018fip, the error on $E_\mathrm{k,SN}$ obtained via the Arnett modelling extends up to the one obtained by the spectral modelling and spherical approximation.}
\label{fig:ks}
\end{center}
\end{figure}

\subsection{The energetics of GRB 180728A/SN 2018fip in context \label{sec:grbsnene}}

The photometric data (Fig. \ref{fig:SNfit}) reveals a light curve that smoothly decays after a peak, then breaks into a steeper decay, and finally gives over to the rising associated supernova SN 2018fip. 
The light curve evolution strongly resembles GRB 011121 \citep{Greiner2003ApJ}. It is less similar to well-known nearby GRB-SN associations like GRB 030329 - SN 2003dh \citep{Stanek2003ApJ,Hjorth2003b,Matheson2003ApJ}, where the afterglow brightness suppresses the SN bump's visibility. Similarly, low-energy events such as 
XRF 060218/SN 2006ap \citep{Campana2006Nature,Pian2006Nature, Ferrero2006AA} and GRB 171205A - SN 2017iuk \citep{Izzo2019a}, show a clear SN bump but 
essentially no afterglow emission.

Figure~\ref{fig:bolocomp} shows the quasi-bolometric light curve of SN~2018fip compared to the 13 GRB-SNe, with low-order spline fits overplotted for clarity. Details regarding the selection criteria for the comparison sample and the methodology used to estimate their bolometric light curves are described in \citet{Kumar2024a}. The comparison sample consists of 13 well-observed GRB-SNe, whose light curves are compiled primarily from \citet{Cano2017a} and references therein, along with more recent events such as SN~2016jca \citep{Cano2017b, Ashall2019a}, SN~2017htp \citep{Postigo2017a, Melandri2019a, Kumar2022a}, SN~2017iuk \citep{Wang2018a, Izzo2019a, Suzuki2019a, Kumar2022a}, and SN~2019jrj \citep{Melandri2022a}. The full sample spans a wide range of peak quasi-bolometric luminosities, from $\sim 3$ to $36 \times 10^{42}$~erg~s$^{-1}$ \citep{Kumar2024a}. SN~2018fip reaches a peak luminosity of $\sim 5.7 \times 10^{42}$~erg~s$^{-1}$ at $\approx 15$ days post-burst, placing it on the fainter end of the GRB-SN luminosity distribution. This highlights the considerable diversity in GRB-SN energetics and positions SN~2018fip as an intrinsically low-luminosity example within the population. Notably, this comparison is conducted without rescaling to SN~1998bw, allowing for a direct interpretation of absolute luminosity differences across the sample.

To further compare the light curves of SN 2018fip with those of other GRB SNe, we used the parameterization k,s\footnote{To minimize errors, we used the weighted mean of k and s obtained from $g'r'z'$-bands light curves following \citep{Klose2019a}.} (top panel of Fig. \ref{fig:ks} and section \ref{sect:lcana}). 
SN 2018fip lies in the middle of the k,s space covered by GRB-SNe. Therefore, it is one of the nearest SNe in the k,s space to 
the prototype SN 1998bw. The most similar case is GRB 140606B, which also has a similar isotropic energy.
The faint and early-peaking end of the diagram (bottom-left) is occupied by the notable cases of SN 2010bh/GRB 100316D \citep[][]{Starling2011a,Bufano2012a,Olivares2015a} and SN 2006aj/GRB 060218 \citep[][]{Ferrero2006AA}. At the luminous/late-peaking end lies 2011kl/GRB111209B, but we note that this event is more similar spectroscopically to superluminous supernovae 
\citep{Greiner2015Nature,Kann2019AA}.
We do not show the BOAT GRB 221009A in the figures because its associated supernova SN 2022xiw cannot be thoroughly studied, although observations suggest that it was less luminous than SN 1998bw 
\citep[e.g.,][]{Blanchard2024a,Fulton2023a,Levan2023a,Kong2024a,Shrestha2023a,Srinivasaragavan2023a}.

In the bottom panel in Figure~\ref{fig:ks} we compare all GRBs with a known SN in the $E_\mathrm{k,SN}$ -- $E_\mathrm{\gamma,iso}$ plane, \citep[data from][thus limited to this about 2021.]{Minaev2020a,Tsvetkova2021a,Demianski2018a,Dainotti2022a}. For SN 2018fip, we also consider the additional contribution of the high-velocity component obtained by spectral modelling, but still in a spherical approximation (section ~\ref{sec:asym}). 
GRB radiated energy (not corrected for collimation) spans more than 6 orders of magnitude, with the recent BOAT being an extreme example \citep{Burns2023a,Frederiks2023a}.
As already known, most GRB-SNe have E$_\mathrm{k,SN}\sim10^{52}$--$10^{53}$ erg
(not corrected for asymmetry, see \citealt{Mazzali2014a, Ashall2019a, Melandri2019a}), regardless of the GRB energy, suggesting that the GRB-SN phenomenon is driven by the SN, not the GRB jet \citep[e.g.,][]{WoosleyBloom2006a,Mazzali2014a}.
SN 2018fip is towards the low-energy side of this plane, below $ 10^{52}$ erg, which means that overall not an extreme amount of energy resulted from the death of this massive star. This could even double if one considers the additional energy of the high-velocity ejecta, although this is likely highly collimated, and thus must be considered an upper limit.
Considering the events at $z<0.2$, it we find it interesting to compare GRB 180728A with the well-studied GRB 030229, again omitting the weakly constrained SN 2022xiw of GRB 221009A.
The top panel shows that SN 2003dh is one of the brightest supernovae, 3 times more luminous, though peaking at similar times as SN 2018fip (see Fig.~\ref{fig:ks}). The bottom panel shows that it has E$_\mathrm{k,SN}\gtrsim5-10$ times that of SN 2018fip.  

\begin{figure}[tp]
\begin{center}
\includegraphics[width=0.5\textwidth,angle=0]{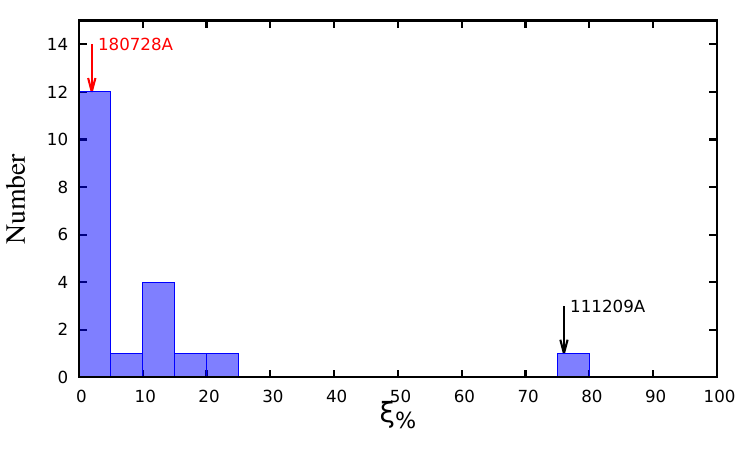}
\caption{Distribution of the GRB efficiency $\xi_{\%}$ 
following the approach described in \citet{Lu2018a}. We highlighted GRBs 180728A/SN~2018fip and 111209A/SN~2011kl that are discussed in the text.}
\label{fig:equi}
\end{center}
\end{figure}

\citet{Lu2018a} have estimated the energy partition within GRB-SNe, defined as the GRB efficiency $\xi_{\%}=E_\mathrm{tot}/(E_\mathrm{tot}+E_\mathrm{k,SN})$\footnote{We prefer to use $\xi_{\%}$ instead of $\eta_{\%}$ like in \citep{Lu2018a}, because it can be confused with the radiative efficiency.}, where $E_\mathrm{tot}$ is the total collimated energy of the jet (section ~\ref{sec:collene}). 
They confirmed previous results showing that in these systems the beaming-corrected GRB energy is usually smaller than the SN energy, with less than 30\% of the total energy distributed in the relativistic jet. 
It is likely that the real distribution is less skewed toward lower values, because in many cases only lower limits exist \citep[for details, see][]{Lu2018a}.
For GRB 180728A/SN 2018fip, using the total collimated energy estimated in section~\ref{sec:collene} and the SN energy found in section~\ref{sec:asym}, we find that the efficiency of GRB 180728A is $\sim2$\%.\footnote{The efficiency varies in the range 1 -- 3 \%, respectively for ISM and wind medium.}
Considering the assumptions made (see sections  \ref{sec:collene} and \ref{sect:SED}), this is a qualitative value that enables comparison with other GRB-SNe.
Similarly to the results obtained by \citet{Lu2018a}, we find that the low efficiency of this event is similar to the majority of GRB-SNe, where only GRB 111209A stands out\footnote{For GRB 111209A, like for the other bursts, we referred to the values reported in \citet{Lu2018a} and the discussion in \citet{Kann2019AA}.} (Fig.\ref{fig:equi}).
Please note again that in these considerations
GRBs are already corrected for the collimation, but the asphericity of SNe is not considered.
In general, for a correct comparison, all events should be
modelled considering a high-velocity component \citep[see, e.g.][]{KhatamiKasen2019a} and also corrected for asphericity, but this is possible only for very few events \citep[e.g.][]{Ashall2019a}.

In the bottom panel of Figure~\ref{fig:ks}, we show in purple all GRBs with a duration shorter than $\sim20$ sec (the average LGRB duration \citep[e.g.,][]{Minaev2020a,Tsvetkova2021a}, and we observe that they typically have $E_\mathrm{\gamma,iso}<10^{53}$ erg. 
Assuming that $E_\mathrm{\gamma,iso}$ is a reasonable good proxy for the total GRB jet energy (or, in other words, assuming universal efficiency),
this can be partly explained in the context of the collapsar scenario. In this model, only the longest GRBs (those with the longest engines) retain sufficient energy after breaking out from the stellar envelope to power the GRB jet \citep[e.g.,][]{Lazzati2013a}. 
Indeed, we notice that, in general, GRB 180728A is in the central part of the figure, in agreement with the aforementioned collapsar scenario and the short duration of just 8.7 s.
Notably, GRB 180728A has one of the shortest durations among GRB-SNe, along with GRB 091127 (7.42 s), which also exhibits a similar afterglow, as noted in Sect.~\ref{sec:agcon}.
GRB 091127 has similar efficiency ($\xi_{\%}\lesssim8$), a slightly more energetic SN ($E_\mathrm{k,SN}\sim13\times10^{51}$ erg), and a very similar afterglow evolution, suggesting that it had a progenitor and engine similar to those of GRB 180728A.
However, its afterglow was four magnitudes more luminous, making the SN-bump less evident \citep{Cobb2010a,Olivares2015a}.

\section{Summary and conclusions}\label{sec:summary_and_conclusions}

GRB 180728A is a long-duration gamma-ray burst (GRB) detected at redshift $z = 0.117$. It is notable for its high isotropic energy release $E_\mathrm{\gamma,iso}>10^{51}$ erg, making it one of the nearest high-energy GRBs ever observed at low redshift (z $<0.2$), similar to GRB 030329. 
We used multi-band photometric and spectroscopic observations up to 80 days post-burst and applied image subtraction to isolate the SN light from a nearby star. This is one of only a handful of events detected in the $J$-band.
We analyzed the prompt and afterglow emission and host-galaxy spectra, and spectral synthesis of the SN was done using TARDIS. 
We then compared GRB 180728A/SN 2018fip with other events to investigate correlations between GRB energy and SN properties. 
In the following, we summarize our key results.
\begin{itemize}
    \item The GRB has an isotropic energy $E_\mathrm{\gamma,iso}$ of $(2.5\pm0.5)\times10^{51}$ erg and an intrinsic spectral peak energy E$_\mathrm{p,i}$ of $123\pm28$ keV. 
    \item The afterglow exhibits a slow early fading followed by a break at 0.2 days. It was slightly under-luminous, placing it between high-energy cosmological GRBs and low-energy nearby GRBs.    
    \item The host is a low-mass, blue, star-forming irregular galaxy, typical for low-z collapsar events. 
    \item SN 2018fip spectral modelling indicates aspherical / asymmetric explosion with a two-component ejecta: a 
    narrow, high-velocity outer layer more dominant during the early spectra 
    (5--8 days in rest-frame) with a velocity above $20,000$ km s$^{-1}$,
    and a slower ($15,000$ km s$^{-1}$), more massive inner spherical component more dominant in later spectra (16--21 days in rest-frame). 
    The high-velocity component is likely confined to $<30^{\circ}$ and plays a crucial role in explaining the evolution of SN 2018fip.
    \item Assuming a quasi-spherical scenario, the spectral modelling yields a total ejected mass of $M_\mathrm{ej}=3\,M_\odot$ and a kinetic energy of $E_\mathrm{k,SN}\sim6.5\times10^{51}$ erg. 
    The symmetric spherical model of Ni-radioactive-heating of \citet{Arnett1982a}, gives $M_{Ni} = 0.19^{+0.15}_{-0.09}$ M$_{\odot}$, $M_\mathrm{ej} = 2.4^{+1.7}_{-0.7}$ M$_{\odot}$, and $E_\mathrm{k,SN} = 3.2^{+2.3}_{-1.0}\times10^{51}$ erg.
    \item SN 2018fip is a a broad-lined Type Ic with a kinetic energy just below the common range of $10^{52}$--$10^{53}$ erg typical for GRB-SNe.

\end{itemize}

In conclusion, despite its high energy, GRB 180728A is associated with a supernova and afterglow that are intrinsically fainter than those of typical events.
Indeed, although its burst energy is comparable to that of GRB 030329, 
both its afterglow and supernova are significantly fainter, 
highlighting the diversity in GRB-SN properties even at similar redshifts and energies.
Overall, the low GRB efficiency of 180728A ($\sim2$\%) is similar to most GRB-SNe, reinforcing the view that these explosions are powered primarily by the supernova rather than the GRB jet. 
Finally, our results reinforce earlier findings on the asphericity of GRB–SN explosions and emphasize its crucial role in shaping the true energetics.
The upcoming extremely large telescopes and the \textit{James Webb} Space Telescope \citep[][]{McGuire2016a} will enable multi-epoch, high-quality spectroscopy of GRB-SNe beyond $z \sim 0.2$,
expanding the sample available for advanced spectral modelling, including studies of asymmetry, and allow investigation of the higher-energy GRBs typical at cosmological distances.

\begin{acknowledgements}

We thank the anonymous referee for providing thoughtful comments. 
A. Rossi thanks P. Evans for his help in the analysis of the {\it Swift}/XRT data.
A. Rossi and E. Palazzi acknowledge financial support from \textit{from PRIN-MIUR 2017 (grant 20179ZF5KS)} and from the INAF project \textit{Supporto Arizona \& Italia}. 
L. Izzo acknowledges financial support from the INAF Data Grant Program 'YES' (PI: Izzo) {\it Multi-wavelength and multi messenger analysis of relativistic supernovae}.
K. Maeda acknowledges support from the Japan Society for the Promotion of Science (JSPS) KAKENHI grants JP24KK0070, JP24H01810, and JP20H00174. The spectral mdoeling calculations were in part carried out on Yukawa-21 at YITP in Kyoto University. 
DBM is funded by the European Union (ERC, HEAVYMETAL, 101071865). Views and opinions expressed are, however, those of the authors only and do not necessarily reflect those of the European Union or the European Research Council. Neither the European Union nor the granting authority can be held responsible for them. The Cosmic Dawn Center (DAWN) is funded by the Danish National Research Foundation under grant DNRF140.
AK is supported by the UK Science and Technology Facilities Council (STFC) Consolidated grant ST/V000853/1.
R.B acknowledges funding from the Italian Space Agency, contract ASI/INAF n. I/004/11/6.
The research leading to these results has received funding from the European Union's Horizon 2020 Programme under the AHEAD2020 project (grant agreement n. 871158).
Part of the funding for GROND (both hardware and
personnel) was generously granted by the Leibniz-Prize to G. Hasinger (DFG grant HA 1850/28-1) and by the Th\"uringer Landessternwarte Tautenburg.
The VL, PB, DV study was conducted under the state assignment of Lomonosov MSU.
This work made use of WISeREP - https://www.wiserep.org.
This research has made use of the NASA/IPAC Infrared Science Archive, which is funded by the National Aeronautics and Space Administration and operated by the California Institute of Technology. 
This work made use of data supplied by the UK \swift{} Science Data Centre at the University of Leicester. 
This research made use of TARDIS, a community-developed software package for spectral synthesis in supernovae. The development of TARDIS received support from GitHub, the Google Summer of Code initiative, and ESA’s Summer of Code in Space program. TARDIS is a fiscally sponsored project of NumFOCUS. TARDIS makes extensive use of Astropy and Pyne.

\end{acknowledgements}

\bibliographystyle{aa}
\bibliography{biblio} 



\begin{appendix} 


\section{Lorentz factor \label{sec:lorentz}}

Under the assumptions of the light curve analysis described in section \ref{sect:lcana} 
we are able to derive the Lorentz factor of the jet. Within the fireball forward shock model the peak of the optical emission corresponds to the afterglow onset
\citep{SariPiran1999a}, which in rest-frame corresponds to the deceleration timescale 
$t_{dec}\sim R_{dec}/(2c\Gamma^{2}_{dec})$, where $R_{dec}$ is the decelarion radius, c the speed of light and $\Gamma_{dec}$ is the Lorentz factor at $t_{dec}$. The initial Lorentz factor $\Gamma_0$ is expected to be twice that of $\Gamma_{dec}$ \citep{PanaitescuKumar2000a,Meszaros2006a}. 
Following \citet{Molinari2007AA} which assumes a homogeneous surrounding medium, we have

\begin{equation}\label{eq:lorentz}
\Gamma(t_{peak})= 160 \left[ \frac{ E_{\gamma,53} \, (1+z)^{3}}{\eta \, n \, t^{3}_{peak,2}}\right]^{1/8}\,,
\end{equation}

where $E_\mathrm{\gamma,iso}$ is the isotropic-equivalent energy released by the GRB in gamma rays
and in units of $10^{53}$ erg, 
$n=1\,\rm cm^{-3}$ is the number density of the constant medium, 
t$_{peak,2}$ = t$_{peak}/$(100 s), 
$\eta$ is the radiative efficiency fixed to 0.2 (see Sect.~\ref{sec:collene}). 
The peak time is t$_\mathrm{peak}=t_\mathrm{b,peak}(-\alpha_\mathrm{r}/\alpha_{d})^{(1/[n(\alpha_\mathrm{d}-\alpha_\mathrm{r})]}$, where $n$, $t_\mathrm{b,peak}$, $\alpha_\mathrm{d}$, $\alpha_\mathrm{r}$ are the smoothness (fixed to 10), the break time and the indexes of the rise and decay phase identified by the broken power-law model (Sect.~\ref{sect:lcana}). 
Using the values in section \ref{sect:lcana}
we find the peak time $t_\mathrm{peak}=219\pm20$ s.
Using $E_{\gamma,53}=2.5\times10^{-2}$ (see Sect.~\ref{sec:grbene}), we obtain $\Gamma_0\approx80 \,(\eta\,n)^{-1/8}$
with typical values $\eta=0.2$ and $n=1\,\rm cm^{-3}$.


\section{Dust extinction \label{sec:dust}}

\begin{figure}[!thp]
\begin{center}
\includegraphics[width=0.5\textwidth,angle=0]{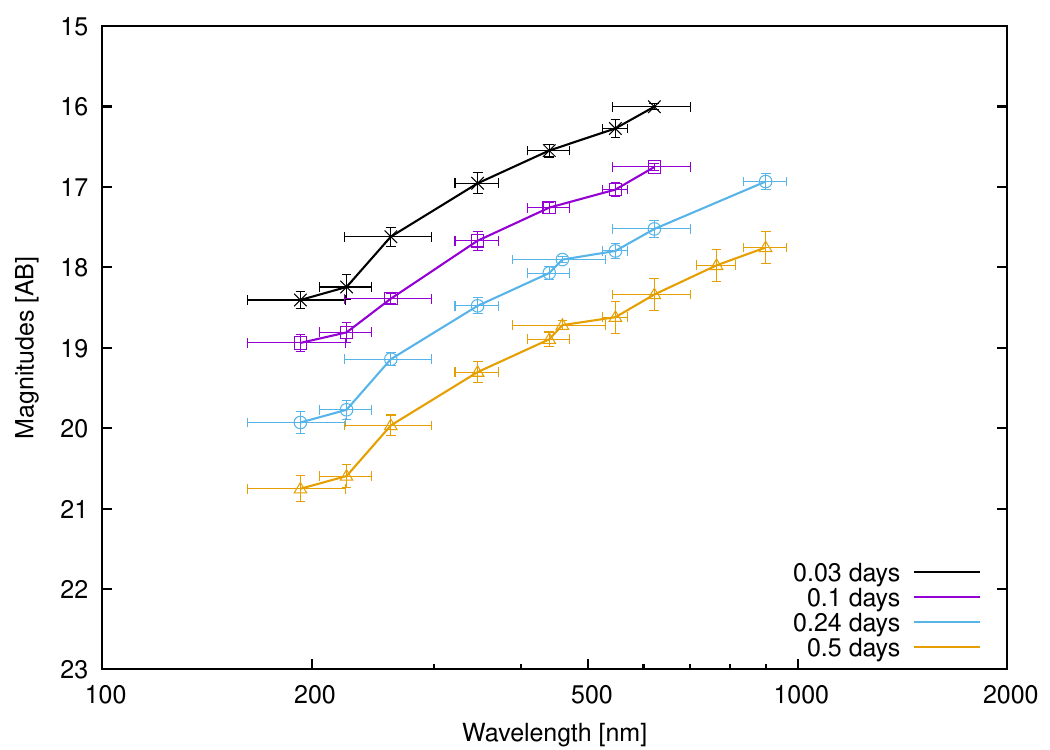}
\caption{UV-optical-NIR SEDs between 0.03 and 0.5 days. 
The Galactic 2175~\AA{} absorption feature is clearly visible in the $uvm2$-band.
}
\label{fig:sedopt}
\end{center}
\end{figure}

In section section \ref{sect:SED}, we find that the optical-to-X-ray SEDs are best modelled with the LMC extinction law.
From Fig.~\ref{fig:sedopt} one can notice that 
the Galactic 2175~\AA{} absorption feature is clearly visible in the $uvm2$-band in early epochs, 
and thus the LMC and MW models are favoured. 
Multi-epoch fits of the SEDs before 1.5 days, using the Milky Way (MW), Large and Small Magellanic Cloud (LMC, SMC) dust attenuation curves of \citep{Pei1992a} give the following results:
\begin{itemize}
\item MW dust: $\beta=0.637\pm0.137$, $A_V=0.023\pm0.095$ mag for $\chi^2$/d.o.f$.=0.73$
\item LMC dust: $\beta=0.570\pm0.201$, $A_V=0.081\pm0.110$ mag for $\chi^2$/d.o.f$.=0.70$
\item SMC dust: $\beta=0.564\pm0.175$, $A_V=0.061\pm0.098$ mag for $\chi^2$/d.o.f$.=0.69$
\end{itemize}
Across all cases, extinction remains negligible within measurement uncertainties, with a modest preference for the LMC model over the MW model.

\section{On the SN modelling \label{sec:addsn}}

\begin{figure}[!htp]
\begin{center}
\includegraphics[width=0.49\textwidth]{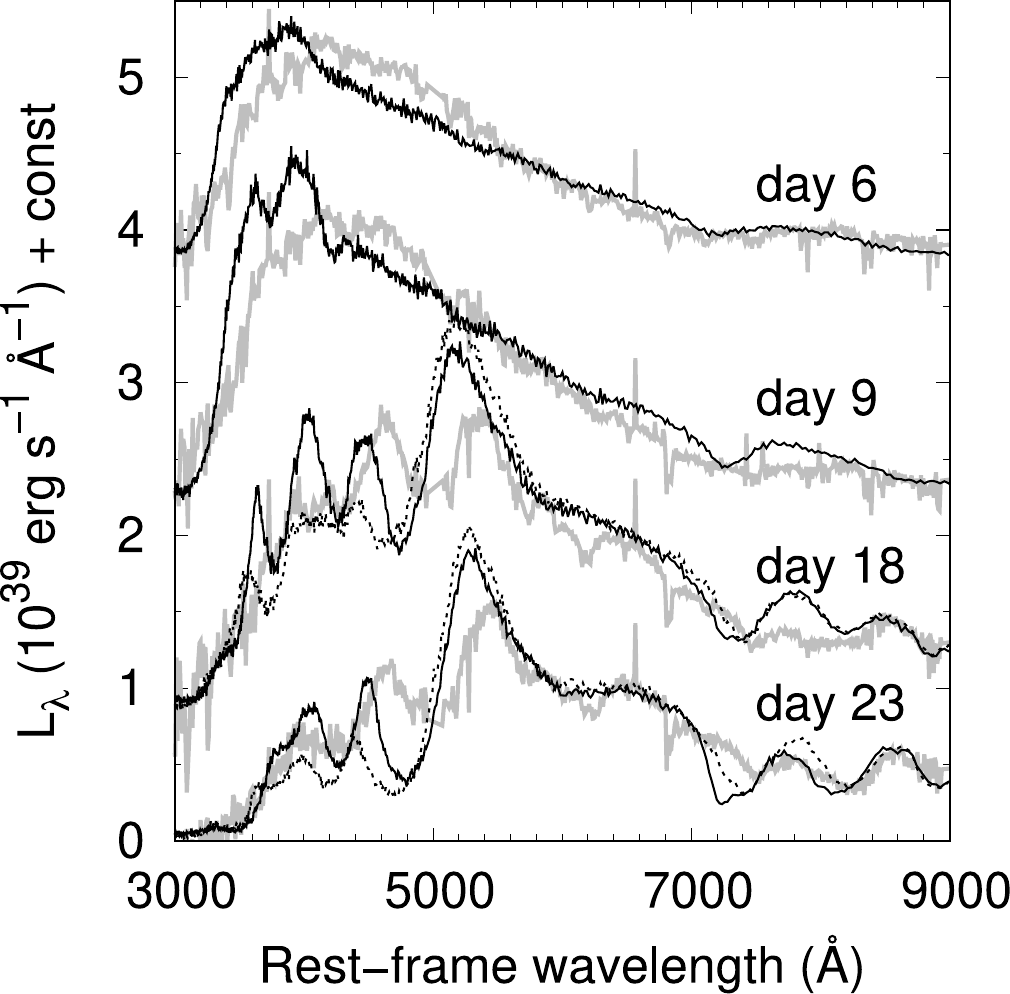}
\caption{Another set of the spectral models that {\em do not} fit to the observed spectra. The solid curves show the synthesized spectra for the model in which the mass fractions of $^{56}$Ni and $^{58}$Ni are decreased by a factor of 8 as compared to the reference model. The dashed curves adopt a pure C+O composition (plus the progenitor metals) for the high-velocity component. The gray lines show the observed spectra (see the caption of Fig. \ref{fig:spec_ref} for details). 
}
\label{fig:spec2}
\end{center}
\end{figure}

In section \ref{sec:snmod} we find that the high-velocity, Fe-rich outer ejecta component that fits early spectra produces clearly incorrect spectral features at later epochs. Here we note
that this problem is essentially independent from the ejecta composition and density in the inner region; it is the existence of the Fe-rich high-velocity component inferred from the earliest phase-spectra that produces too much absorption even in the later epochs, irrespective of the nature of the input radiation that irradiates the outer region. We have varied both the inner and outer structure and composition in several ways, but have not found a solution that can explain the early and late-time spectra simultaneously. As an example, Fig. \ref{fig:spec2} shows the synthetic spectra for a model in which X($^{56}$Ni) and X($^{58}$Ni) in the outer, high-velocity ($> 20,000$ km s$^{-1}$) component are decreased by a factor of 8 from our reference model (i.e., the same one used for the region in $15,000 - 20,000$ km s$^{-1}$). While the absorption below $\sim 4,000$\AA\ is not sufficient on day 6 and 9, the model already shows the clear problems for the spectra on day 18 and 23. Indeed, the density in the outer region obtained for day 6 and 9 is too large even if this is dominated by the C+O material to be compatible to the spectra on day 18 and 23.
As yet another exercise for the spectral analysis on day 18 and 23, we have added pure C+O material (plus the progenitor metals) above $20,000$ km s$^{-1}$ and changed the slope of the density structure as shown by the dashed line in Fig. \ref{fig:ejecta}. 
We consider this to be an upper limit necessary to not produce too strong high-velocity O I and NIR Ca II, and in fact it is below the one inferred from the spectral analysis on days 6 and 9.

\begin{figure}[!htp]
\begin{center}
\includegraphics[width=0.49\textwidth,angle=0]{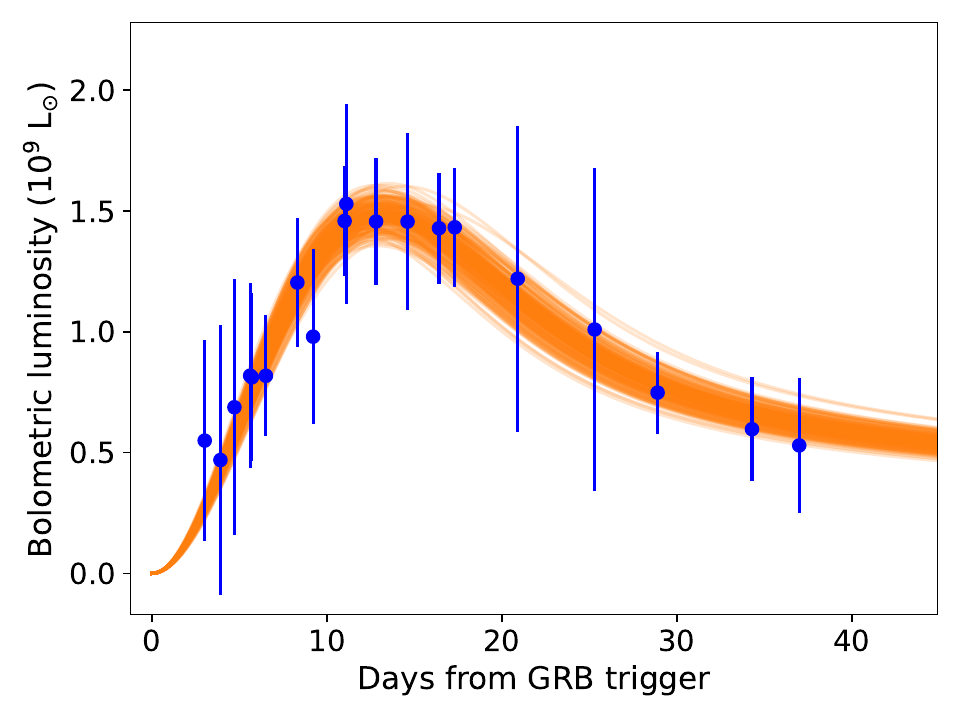}
\caption{Monte Carlo Markov Chain fit to the bolometric light curve of SN 2018fip, obtained from our $g'r'i'z'J$ photometry, using the radioactive-heating mode \citep{Arnett1982a}.
}
\label{fig:bolo}
\end{center}
\end{figure}

\begin{table}
\centering
\caption{Reference stars used for MASTER photometry.}
\begin{tabular}{lll} 
\toprule
Gaia DR2 source id  & RA Dec (J2000) &  Gaia $g$ mag \\
\midrule
5929817258354020000 & 253.5424 $-$54.0430 & 14.781\\
5929817395792950000 & 253.5914 $-$54.0381 & 14.078\\
5929817430152690000 & 253.6073 $-$54.0314 & 13.581\\
5929817361433580000 & 253.5256 $-$54.0165 & 14.980\\
5929817185290700000 & 253.5025 $-$54.0461 & 14.200\\
5929820178931820000 & 253.4949 $-$54.0292 & 14.685\\
5929796642509320000 & 253.4683 $-$54.0649 & 14.025\\
5929816571159180000 & 253.6535 $-$54.0655 & 13.114\\
5929818014268200000 & 253.6909 $-$54.0304 & 14.202\\
\bottomrule
\end{tabular}
\tablefoot{
In addition to these reference stars we selected a large list of comparison stars with similar brightness to the object. This was done to determine the measurement error of its magnitude as a luminosity variation of these stars 
\citep[See][for more details]{Lipunov2019a}.
}
\label{tab:master-stars}
\end{table}

\begin{table}
\centering
\caption{Reference stars for GROND and X-shooter photometry.}
\begingroup
\setlength{\tabcolsep}{4pt} 
\footnotesize
\begin{tabular}{lcccc} 
\toprule
RA Dec (J2000) &  g$^\prime$ & r$^\prime$ & i$^\prime$ & z$^\prime$   \\
\midrule
253.5730    $-$54.0389 & 16.728 (1) & 16.027  (1)  & 15.750  (1)  & 15.559  (1)  \\ 
253.5714    $-$54.0442 & 17.200 (2) & 16.635  (1)  & 16.410  (2)  & 16.260  (2)  \\ 
253.5738    $-$54.0550 & 17.425 (2) & 16.758  (1)  & 16.518  (2)  & 16.355  (2)  \\ 
253.5807    $-$54.0370 & 18.022 (3) & 17.331  (2)  & 17.076  (3)  & 16.903  (5)  \\ 
253.5661    $-$54.0404 & 18.072 (5) & 17.466  (3)  & 17.220  (6)  & 17.044  (6)  \\ 
253.5681    $-$54.0489 & 18.277 (3) & 17.558  (2)  & 17.225  (3)  & 17.014  (3)  \\ 
253.5900    $-$54.0397 & 18.326 (5) & 17.571  (2)  & 17.278  (5)  & 17.078  (5)  \\ 
253.5509    $-$54.0445 & 18.321 (5) & 17.582  (2)  & 17.282  (3)  & 17.077  (5)  \\ 
253.5566    $-$54.0514 & 18.449 (5) & 17.801  (2)  & 17.550  (5)  & 17.392  (5)  \\ 
253.5591    $-$54.0291 & 18.563 (7) & 17.851  (5)  & 17.542  (6)  & 17.340  (7)  \\ 
253.5573    $-$54.0339 & 18.958 (8) & 18.014  (3)  & 17.635  (5)  & 17.397  (6)  \\ 
253.5641    $-$54.0485 & 18.901 (6) & 18.114  (3)  & 17.825  (6)  &  ---   \\ 
253.5727    $-$54.0487 & 18.853 (6) & 18.173  (5)  & 17.888  (7)  &  ---   \\ 
253.5641    $-$54.0506 & 19.125 (8) & 18.311  (5)  & 17.924  (7)  &  ---   \\ 
253.5662    $-$54.0461 & 19.323 (8) & 18.638  (5)  & 18.329  (8)  &  ---   \\ 
253.5536    $-$54.0400 & 19.522 (9) & 18.699  (6)  & 18.387  (9)  &  ---   \\ 
253.5591    $-$54.0473 & 19.584 (9) & 18.856  (7)  &  ---   &  ---   \\ 
253.5602    $-$54.0376 & ---  & 19.243  (7)  &  ---   &  ---   \\ 
253.5779    $-$54.0431 & ---  & 19.466  (9)  &  ---   &  ---   \\ 
\bottomrule
\end{tabular}
\tablefoot{%
AB magnitudes obtained using a $2\times\mathrm{FWHM}$ aperture.
They have been obtained using zeropoints calibrated via the observations of the STD17 SDSS field.
Numbers in parentheses give the photometric 1$\sigma$ statistical uncertainty of the secondary standards in units of  milli-mag.}
\label{tab:stdgrond}
\endgroup

\end{table}

\onecolumn
\begin{longtable}{lcrr}
\caption{\label{tab:photalt} Observations of the afterglow of GRB 180728A.}\\
\toprule
$\Delta$t & Brightness & Filter & Telescope/Instrument \\
 (day) & (AB Mag) &  &  \\
\toprule
0.028778	& $	18.48	\pm	0.19			$ & $	uvw2	$ &	\emph{Swift} UVOT	\\
0.034915	& $	18.55	\pm	0.10			$ & $	uvw2	$ &	\emph{Swift} UVOT	\\
0.092967	& $	18.74	\pm	0.11			$ & $	uvw2	$ &	\emph{Swift} UVOT	\\
0.356316	& $	20.5	\pm	0.13			$ & $	uvw2	$ &	\emph{Swift} UVOT	\\
0.495682	& $	20.83	\pm	0.16			$ & $	uvw2	$ &	\emph{Swift} UVOT	\\
0.818226	& $	21.65	\pm	0.18			$ & $	uvw2	$ &	\emph{Swift} UVOT	\\
1.027922	& $	21.96	\pm	0.32			$ & $	uvw2	$ &	\emph{Swift} UVOT	\\
3.484493	& $ >	22.77					$ & $	uvw2	$ &	\emph{Swift} UVOT	\\
6.975032	& $ >	23.25					$ & $	uvw2	$ &	\emph{Swift} UVOT	\\
13.650938	& $ >	23.66					$ & $	uvw2	$ &	\emph{Swift} UVOT	\\
22.073895	& $ >	23.08					$ & $	uvw2	$ &	\emph{Swift} UVOT	\\
\midrule				
...	& 					 & 	&		\\
\midrule
3.329745	& $	19.88	\pm	0.34			$ & $	K_{s}	$ &	2.2m MPG/GROND	\\
4.366551	& $ >	19.7					$ & $	K_{s}	$ &	2.2m MPG/GROND	\\
5.279745	& $ >	19.8					$ & $	K_{s}	$ &	2.2m MPG/GROND	\\
6.336574	& $ >	19.8					$ & $	K_{s}	$ &	2.2m MPG/GROND	\\
7.247569	& $ >	19.1					$ & $	K_{s}	$ &	2.2m MPG/GROND	\\
10.284028	& $ >	19.0					$ & $	K_{s}	$ &	2.2m MPG/GROND	\\
12.392361	& $ >	19.7					$ & $	K_{s}	$ &	2.2m MPG/GROND	\\
14.351852	& $ >	19.5					$ & $	K_{s}	$ &	2.2m MPG/GROND	\\
16.320602	& $ >	19.6					$ & $	K_{s}	$ &	2.2m MPG/GROND	\\
19.271991	& $ >	19.1					$ & $	K_{s}	$ &	2.2m MPG/GROND	\\
23.343750	& $ >	19.6					$ & $	K_{s}	$ &	2.2m MPG/GROND	\\
28.299769	& $ >	19.5					$ & $	K_{s}	$ &	2.2m MPG/GROND	\\
32.302083	& $ >	19.3					$ & $	K_{s}	$ &	2.2m MPG/GROND	\\
38.319444	& $ >	19.1					$ & $	K_{s}	$ &	2.2m MPG/GROND	\\
47.292824	& $ >	19.6					$ & $	K_{s}	$ &	2.2m MPG/GROND	\\
88.304398	& $ >	19.1					$ & $	K_{s}	$ &	2.2m MPG/GROND	\\
\bottomrule
\end{longtable}
\tablefoot{ The final table will be made available as online material.
All data are in AB magnitudes and not corrected for Galactic foreground extinction.
X-Shooter, UVOT-$ubv$ and $white$ filters, and GROND data result from subtracting the constant emission from the host and the nearby star, while the remaining UVOT bands, REM and MASTER data are not affected noticeably by additional components.
Midtimes are derived 
with the geometric mean of start and stop times: 
$t=\sqrt{(t_1-t_0)\times(t_2-t_0)}$, hereby $t_{1.2}$ are the absolute start and stop times, and $t_0$ is the \emph{Swift} trigger time.}\\


\end{appendix}

\end{document}